\documentclass[journal]{IEEEtran}

\usepackage{xcolor,soul,framed} 

\colorlet{shadecolor}{yellow}
\usepackage[pdftex]{graphicx}
\graphicspath{{../pdf/}{../jpeg/}}
\DeclareGraphicsExtensions{.pdf,.jpeg,.png}

\usepackage[cmex10]{amsmath}
\usepackage{array}
\usepackage{mdwmath}
\usepackage{mdwtab}
\usepackage{eqparbox}
\usepackage{url}
\usepackage{commath}
\usepackage{amssymb}
\usepackage{comment}
\usepackage{amsmath}
\usepackage{cite}
\usepackage{placeins}
\usepackage{float}
\usepackage{makecell}
\usepackage{mathtools}
\usepackage{multirow}
\usepackage{graphicx}

\usepackage{threeparttable}
\usepackage{adjustbox}
\usepackage{tikz}
\usepackage{pdfpages} 
\usepackage{pgffor} 

\makeatletter
\AtBeginDocument{\let\LS@rot\@undefined}
\makeatother

\def\supplementfilename{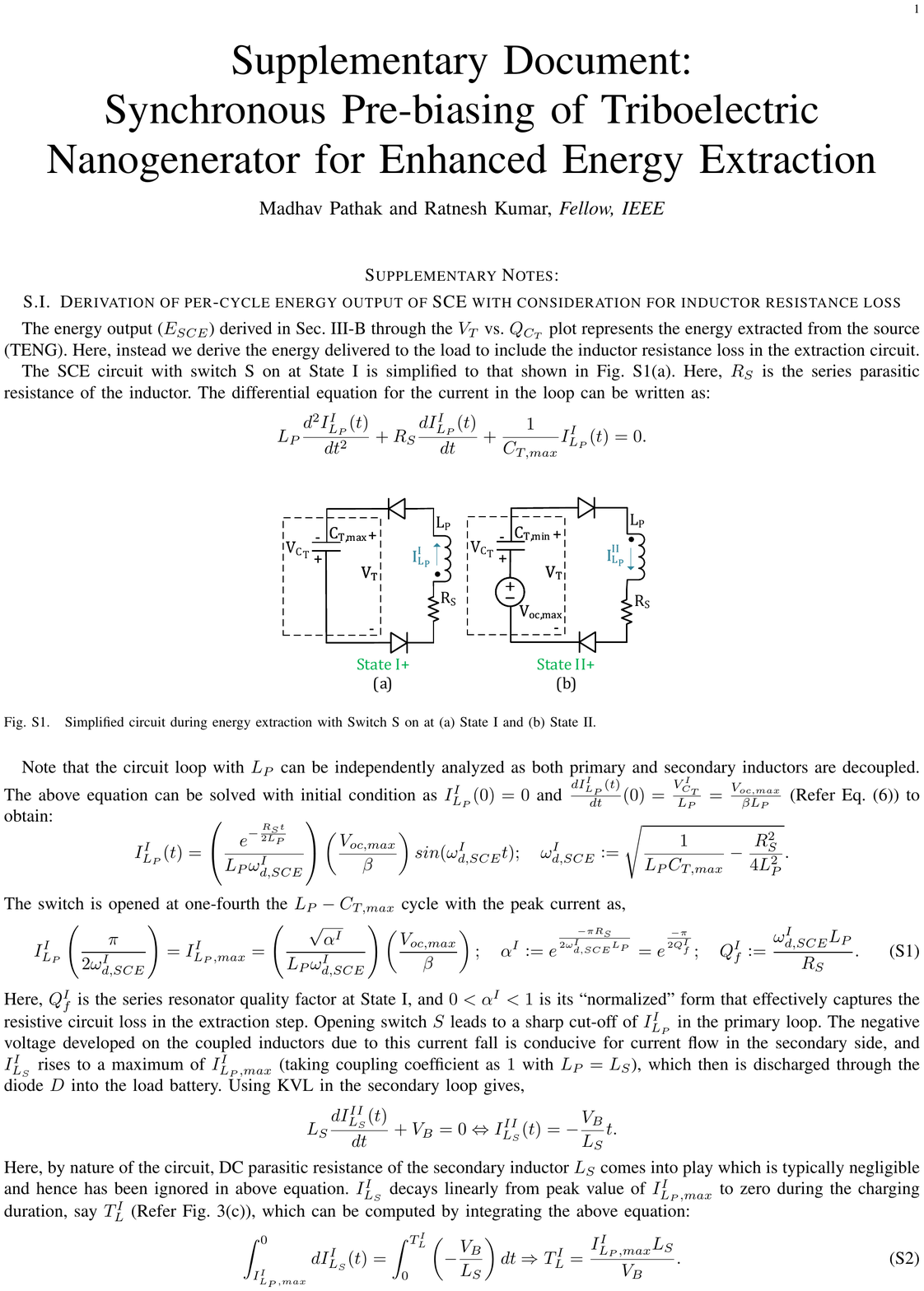}

\pdfximage{\supplementfilename}
\def\numbersupplementpages{\the\pdflastximagepages}

\newif\ifarXiv
\arXivtrue 

\newcommand\copyrighttext{%
  \footnotesize This work has been submitted to the IEEE for possible publication. Copyright may be transferred without notice, after which this version may no longer be accessible.}
\newcommand\copyrightnotice{%
\begin{tikzpicture}[remember picture,overlay]
\node[anchor=south,yshift=10pt] at (current page.south) {\fbox{\parbox{\dimexpr\textwidth-\fboxsep-\fboxrule\relax}{\copyrighttext}}};
\end{tikzpicture}%
}

\usepackage{xr}

\makeatletter
\newcommand*{\addFileDependency}[1]{
  \typeout{(#1)}
  \@addtofilelist{#1}
  \IfFileExists{#1}{}{\typeout{No file #1.}}
}
\makeatother

\newcommand*{\myexternaldocument}[1]{%
    \externaldocument{#1}%
    \addFileDependency{#1.tex}%
    \addFileDependency{#1.aux}%
}

\myexternaldocument{supplement}

\newtheorem{r1}{Remark}
\begin{document}
\bstctlcite{IEEEexample:BSTcontrol}
    \title{Synchronous Pre-biasing of Triboelectric Nanogenerator for Enhanced Energy Extraction}
  \author{Madhav~Pathak and~Ratnesh~Kumar,~\IEEEmembership{Fellow,~IEEE}

\thanks{This work was supported in part by U.S. National Science Foundation under grants NSF-CCF-1331390, NSF-ECCS-1509420, NSF-PFI-1602089, and NSF-CSSI-2004766. The authors would like to thank Alex Denny of Rowe Electronics for PCB implementation of the circuits.}

  \thanks{The authors are with the Department of Electrical and Computer Engineering, Iowa State University, Ames, IA, 50011 USA (e-mail: mpathak@iastate.edu; rkumar@iastate.edu).}
  \vspace*{-.3in}
 }  


\maketitle

\copyrightnotice
\begin{abstract}
Triboelectric Nanogenerator (TENG) is a class of ambient mechanical energy harvesters used to augment the battery life of electronic devices such as sensors in implantables, wearables, and Internet of Things (IoT) applications. In this work, the fundamentals of pre-biasing (pre-charging) the TENG at the start of the operation cycle to enhance the per-cycle extracted energy is presented. The energy gain is mathematically formulated, and the optimum pre-biasing voltage (equivalently charge) is derived by analyzing the energy exchange between the mechanical and the electrical domain over a periodic cycle. Further, a novel Energy Extraction Circuit (EEC) termed as ``Pre-biased Synchronous Charge Extraction (pSCE)" is introduced to 1) Realize synchronous pre-biasing of TENG using the load battery itself and 2) Achieve enhanced energy extraction from TENG. Energy output per-cycle is derived analytically for the pSCE circuit and compared to the state of the art Synchronous Charge Extraction (SCE) circuit. The experimental implementation is performed for the proposed pSCE circuit that shows a 6.65 fold gain over the Full Wave Rectifier (standard EEC) and 1.45 over the SCE circuit for a 5V battery load.  
\end{abstract}


%
\IEEEpeerreviewmaketitle


\vspace*{-.1in}
\section{Introduction} \label{Sec:Intro}
\begin{figure*}
\vspace*{-.1in}
  \begin{center}
  \includegraphics[width=1.0\linewidth]{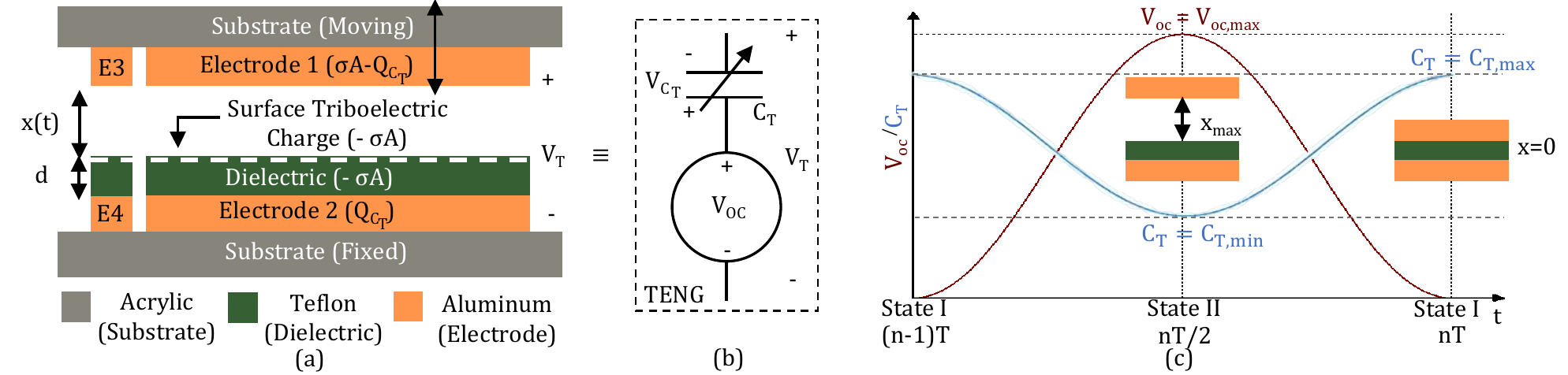}
  \vspace*{-.25in}
  \caption{(a) Cross section view of a contact-separation TENG with a parallel auxiliary TENG(b) Equivalent circuit model of TENG (c) Periodic variation of open circuit voltage ($V_{oc}(t)$) and TENG capacitance ($C_T(t)$) over a period of time $T$.}\label{StripI}
  \end{center}
\vspace*{-.35in}
\end{figure*}
\IEEEPARstart{W}{ireless} sensor nodes used in the Internet of Things (IoT), wearable and, implantable applications are commonly powered by an on-board battery. 
Embedding an ambient energy harvester to charge the on-board battery is a green solution to prolonging the battery life\cite{shaikh2016energy}, and it may also help reduce the size and weight by using a lower capacity battery. 
Among the ambient energy sources such as solar, mechanical, chemical, and thermal\cite{TUNA201625}, mechanical sources are most readily available in the form of wind/water flow, machine/structure vibration, human body motion, etc.\cite{mohanty2019vibration,cai2020recent,wang2020state}. Hence, mechanical to electrical energy transducers such as piezoelectric, electrostatic, and electromagnetic have seen significant development efforts in the past years\cite{maamer2019review,boisseau2012electrostatic}. The piezoelectric energy harvesters have been widely used but are subject to a limited choice of materials and require complex fabrication techniques\cite{yang2018high}. In contrast, Triboelectric Nanogenerator (TENG) has shown promising potential 
owing to its almost universal choice of materials and operation modes\cite{wu2019triboelectric}. 


Motion transduction leads to variable output that needs rectification to charge DC battery load, and thus, Full Wave Rectifier (FWR) can be considered the simplest Energy Extraction Circuit (EEC)\cite{niu2015optimization}. Approaches to enhance TENG's energy output 
require improving the source to load matching. 
To this end, several novel EEC architectures have been proposed in the literature to increase the energy output beyond the FWR circuit such as \cite{zi2015standards,zi2017inductor,zi2016effective,cheng2017high, xi2017universal, cheng2019power,rawy2020triboelectric,wu2020high}. It is shown in \cite{zi2015standards} that all these different EEC outputs are confined within the cycle for maximized energy output (CMEO), 
where the {\em synchronously serial switched flyback converter with the rectified TENG output} can achieve the CMEO irrespective of the load value\cite{cheng2017high,cheng2019power,wu2020high}. This last architecture has been proposed previously in the piezoelectric context\cite{lefeuvre2005piezoelectric,dicken2012power,singh2015broadband,singh2018self}, by the name of Synchronous Charge Extraction (SCE), which is how we also refer to it in this paper.


The aforementioned EECs may be viewed as {\em passive EECs}, where no harvested energy is fed back to the TENG. In this work, we study the {\em pre-biasing} or {\em pre-charging} of TENG by feeding back some of the load battery energy to the TENG, thereby realizing an {\em active EEC}, which enables us to achieve a {\em net output beyond the CMEO limit}. Pre-biasing is accomplished by charging the TENG capacitor at the start of each half-cycle (at the minimum and maximum separation of TENG plates) by using the battery to charge the TENG capacitor (termed pre-biasing) so that it increases the force between the two plates and thereby the work done against it during separation phase by the mechanical source, causing a higher level of transduced energy and hence also a higher level of extracted energy. In contrast, during the contraction phase, when plates move in the direction of the electrostatic force, there is energy loss, and pre-biasing is used to minimize this loss to zero. 
Upper and optimum limits, if any, on the pre-biasing voltage for each half-cycle are derived by analyzing the interplay of mechanical motion and electrostatic forces. 
The concept of pSCE circuit itself was initially explored in our earlier conference paper\cite{pathak2019pre} but without a detailed analysis and experimental validation. Here we present its complete mathematical working and derive the energy output per-cycle, which is then contrasted against the SCE circuit's CMEO output. Additionally, the precise conditions on the pre-biasing voltages for ensuring net benefit over SCE are derived. Finally, the experimental implementation of the pSCE circuit is also presented, and its performance improvement over SCE and the standard Full Wave Rectifier (FWR) circuit is experimentally validated.

It is further important to note that while pSCE circuit architecture has been explored previously for piezoelectric transducers \cite{elliott2012implementation,dicken2012power}, the analysis is entirely different in the case of triboelectric transducers, since it has different circuit model: For a piezoelectric transducer, capacitor appears in {\em parallel} to the source and is {\em fixed}, whereas the triboelectric capacitor appears in {\em series} and is {\em time-varying}. The work reported here involves innovation to deal with the time-varying nature, namely, a “smart discretization” that performs analysis at the two extremities (plates fully contracted versus separate); {\em no such discretization is needed in the piezoelectric case} as the same set of time-invariant equations remain valid at all instances. The triboelectric analysis framework is thus much more intricate and has been developed in the presented generality for the first time.  Also, owing to the above differences in the circuit model, {\em the switching control circuits of the piezoelectric setting} do not work in the TENG setting, and new switching circuits have also been devised for the first time. 

Also, it should be stated that while we perform pre-biasing using the load battery itself, this can also be achieved by feeding back a fraction of the output charge \cite{liu2019integrated,ghaffarinejad2018conditioning,xia2020inductor} or also by the use of LC circuit oscillation, i.e., by using a parallel or series synchronous switched harvesting on inductor (P-SSHI/S-SSHI) circuit as EECs, as presented in our earlier works \cite{pathak2018modeling,pathak2021synchronous} and in \cite{xu2019boost,li2019sshi,kara202070}.
\vspace{-0.1in}
\section{Preliminaries: TENG Model}
\vspace{-0.05in}
The most generic form of TENG operation is in contact-separation mode, where as depicted in Fig.~\ref{StripI}(a), the two parts of the TENG, namely, a metal film acting as Electrode 1 (upper Al plate here) and a metal film covered with dielectric acting as Electrode 2 (bottom Al plate covered with a Teflon tape) repeatedly comes in contact and separates under the influence of an external motion. Also, as shown in Fig.~\ref{StripI}(a), we have created a separate parallel TENG with a minuscule area compared to the main TENG, termed Aux-TENG, which is tapped through Electrodes 3 and 4 to drive the control circuit of the EEC in synchrony with the main TENG, as detailed in the experimental implementation section (Sec.~\ref{Implementation}). Repeated contact-separation between the two plates generates equal and opposite triboelectric static charges on the two plates, with density, denoted $\sigma$. 
Ignoring the fringing field effect, the resulting field is given by, 
\vspace{-0.025in}
\begin{equation}
\label{E'}
\Vec{E_{\sigma}} = \frac{-\sigma}{\epsilon_{0}}\vec{1}_x,
\vspace{-0.025in}
\end{equation}
where $\epsilon_0$ is the electrical permittivity of air, and $\vec 1_x$ is the unit vector pointing upwards in the direction of increasing $x$. The voltage induced between the two electrodes by the above electric field is termed as open circuit voltage and is given by,
\vspace{-0.025in}
\begin{equation}
V_{oc}(t) = -\int_{0}^{x(t)} \Vec{E_{\sigma}}.\vec 1_{x} dx = \frac{\sigma x(t)}{\epsilon_{0}}.
\vspace{-0.025in}
\end{equation}
Also, TENG forms a variable capacitor with air gap, $x(t)$ and dielectric of thickness $d$ between the two electrodes:
\[C_T(t)=\frac{\epsilon_{0} A}{x(t)+d_{eff}}; \quad d_{eff}=\frac{d}{\epsilon_{d}}.\]
Here, $A$ is the contact surface area of the plates, and $\epsilon_d$ is the relative permittivity of the dielectric layer, making its effective thickness lower by that factor: $d_{eff}=\frac{d}{\epsilon_d}$.

When TENG is connected to an external circuit, the movement of the conduction charge (in the form of free electrons on the electrodes that are separate from the triboelectric charge) from one electrode to another, say $Q_{C_T}(t)$ (Refer Fig.~\ref{StripI}(a)), leads to a capacitor voltage, $V_{C_T}(t)=\frac{Q_{C_T}(t)}{C_T(t)}$ between the two electrodes. Thus, the net TENG voltage ($V_T$) is the superposition of open circuit voltage ($V_{oc}$), owing to the triboelectric charge and its induced electrostatic emf, and the TENG capacitor voltage ($V_{C_T}$), owing to the movement of free conducting charge from one electrode to the other via the pathway of an external circuit:
\vspace{-0.025in}
\begin{equation}\label{VT}
    V_T(t)=V_{oc}(t)-V_{C_T}(t) = \frac{\sigma x(t)}{\epsilon_{0}}-\frac{Q_{C_T}(t)}{C_{T}(t)}.
\end{equation}
\vspace{-0.025in}
The above TENG operating equation leads to the circuit model shown in Fig.~\ref{StripI}(b)) with a variable voltage source ($V_{oc}(t)$) in series with a variable capacitor ($C_T(t)$)\cite{niu_zhou_wang_liu_lin_bando_wang_2014}. 

As visualized in Fig.~\ref{StripI}(c), State I designates one extremity of TENG operation, where the two plates are in contact ($x(t)=0$), resulting in zero open circuit voltage, $V_{oc}(t)=0$, and a maximum capacitance ($C_T(t)=C_{T,max}=\frac{\epsilon_{0} A}{d_{eff}}$). While State II designates the other extremity of the TENG operation, where the two plates are maximally apart ($x(t)=x_{max}$) leading to a maximum open circuit voltage, ($V_{oc}(t)=V_{oc,max}=\frac{\sigma x_{max}}{\epsilon_{0}}$ and minimum capacitance ($C_T(t)=C_{T,min}=\frac{\epsilon_{0} A}{x_{max}+d_{eff}}$). Further for analysis convenience, a system constant, ratio of maximum to minimum capacitance, is introduced: $\beta=\frac{C_{T,max}}{C_{T,min}}=\frac{x_{max+d_{eff}}}{d_{eff}}=\frac{x_{max}}{d_{eff}}+1$.
\vspace{-0.05in}
\section{Synchronous Charge Extraction (SCE)}\label{Sec_SCE}
\begin{figure}
\vspace*{-.1in}
  \begin{center}
  \includegraphics[width=1.0\linewidth]{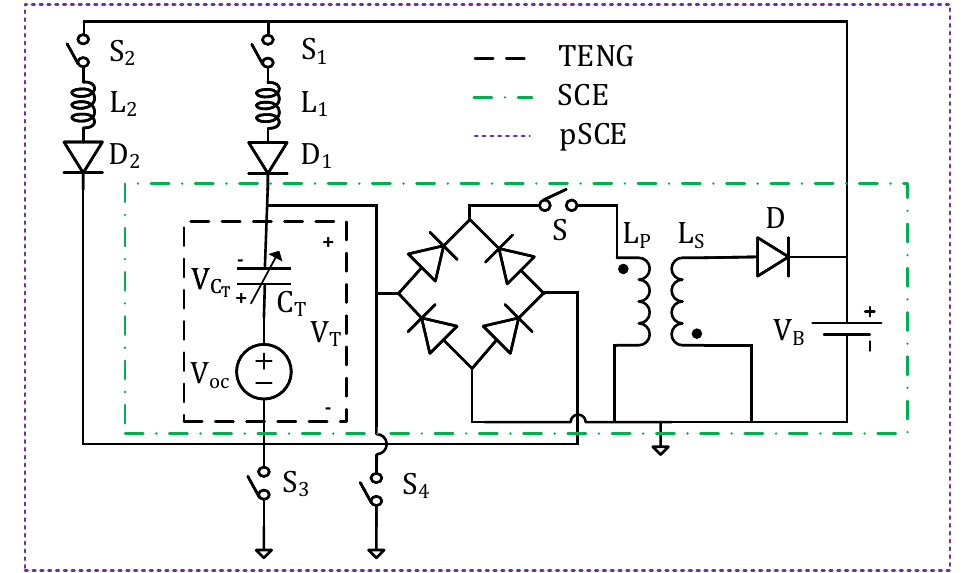}
  \vspace{-0.2in}
  \caption{SCE and pSCE circuit diagram (one inside green box is SCE portion of pSCE which is inside the outer purple box)}\label{Circuit_Diagram}
  \end{center}
\vspace*{-.35in}
\end{figure}

\vspace{-0.05in}
The circuit in the green box of Fig.~\ref{Circuit_Diagram} represents the SCE circuit obtained by cascading a full wave rectifier with the flyback converter \cite{cheng2017high,cheng2019power,wu2020high} via the switch S. 
With switch S open, TENG achieves its high open circuit maximum voltage ($V_{oc,max}$). An efficient energy transfer is made possible by closing the switch S at the extremes (States I and II), thereby forming a $L_P$-$C_T$ oscillator and enabling the transfer of $C_T$ energy to $L_P$ (in one-quarter of the oscillation cycle), and subsequently transferring that energy to the load through $L_s$. 

Here we start by providing a complete analysis of the SCE circuit that derives the TENG voltage and charge at the extremes (States I and II) and use those to derive the per-cycle energy extracted, confirming that it equals the CMEO. This then serves as a baseline for comparing the achieved gain in the extracted energy when pre-biasing is incorporated.
\vspace{-0.15in}
\subsection{Circuit operation}
\begin{figure*}
\vspace*{-.1in}
  \begin{center}
  \includegraphics[width=0.95\linewidth]{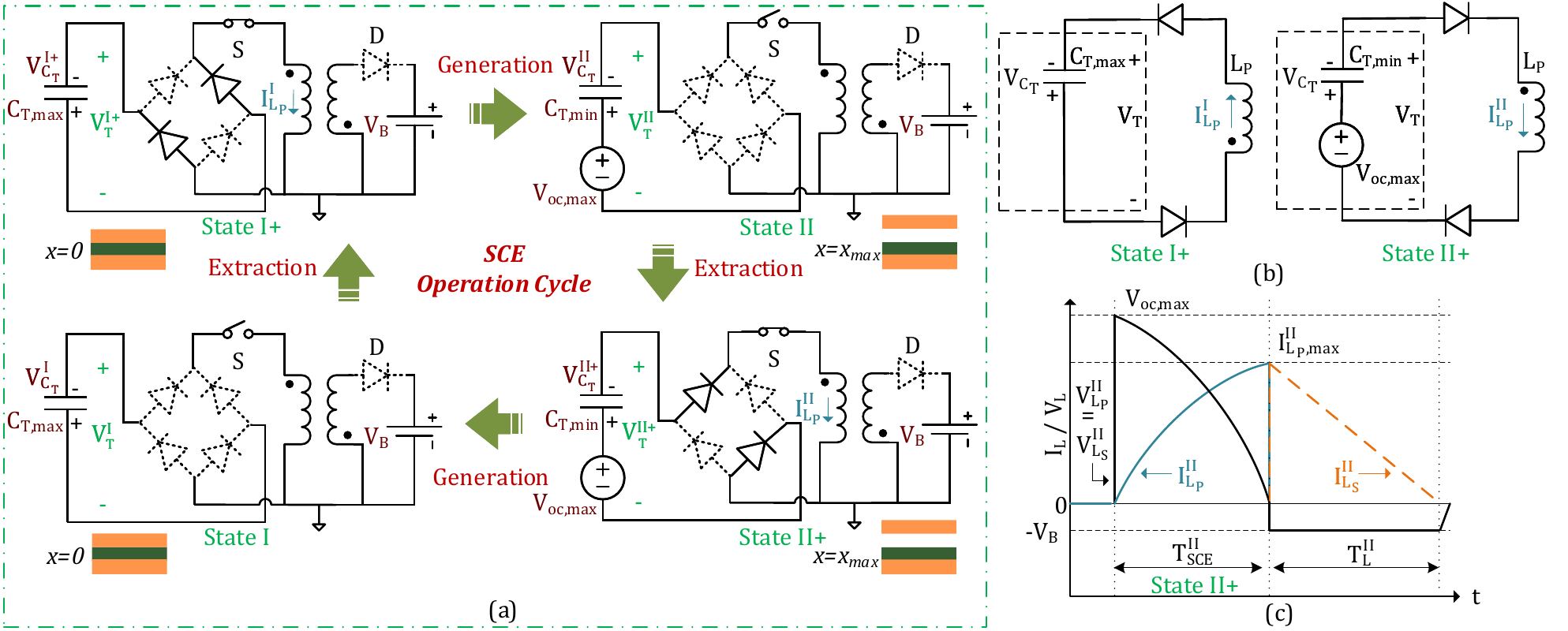}
  \vspace*{-.05in}
   \caption{SCE Circuit (a) Operation cycle, (b) Simplified circuit during energy extraction with Switch S on at States I and II and (c) Primary Inductor Current ($I_{L_P}$), Voltage ($V_{L_P}$) and, Secondary Inductor Current ($I_{L_S}$) during switching at State II.}\label{SCE_Strip}
  \end{center}
\vspace*{-.25in}
\end{figure*}
The SCE circuit operation can be understood by following the operation cycle diagram of Fig.~\ref{SCE_Strip}(a) and TENG voltage waveform of Fig.~\ref{V_Q_Plot}(a). The cycle starts at State I+ with the two plates pressed together ($x=0$) with TENG voltage ($V_T$) and TENG capacitor charge ($Q_{C_T}$) being zero (left top of Fig.~\ref{SCE_Strip}(a)). The switch S is opened and as the two plates are separated, TENG voltage increases with TENG operating in open circuit condition. When the plates are maximally apart, $V_T$ reaches the maxima, with $Q_{C_T}$ still being zero. This is State II (right top of Fig.~\ref{SCE_Strip}(a)), with:
\vspace{-0.025in}
\begin{equation}
    \label{eq:S1}
    V^{II}_{T}=V_{oc,max} ;\quad Q^{II}_{C_T}=0\Rightarrow V_{C_T}^{II}=0. \nonumber
\end{equation}
\vspace{-0.025in}
At this point, switch S is closed resulting in State II+ (right bottom of Fig.~\ref{SCE_Strip}(a)) and simplifying the circuit to as shown in the right half of Fig.~\ref{SCE_Strip}(b). From this figure it can be seen that it is simply a $L_p$-$C_{T,min}$ oscillator, so that the differential equation evolution of the TENG 
capacitor voltage satisfies:
\vspace{-0.025in}
\begin{equation}
\label{eq:D1}
\frac{d^2V_{C_T}(t)}{dt^2}\!+\!\frac{V_{C_T}(t)}{L_PC_{T,min}}\!-\!\frac{V_{oc,max}}{L_PC_{T,min}}\!=\!0; t\!\geq 0; V_{C_T}(0)\!=\!0. 
\end{equation}
\vspace{-0.025in}
Above can be solved for the TENG capacitor voltage $V_{C_T}(t)$ with initial condition as $V_{C_T}(0)=V_{C_T}^{II}=0$ to obtain:
\vspace{-0.025in}
\begin{equation}
V_{C_T}(t)=-V_{oc,max}\cos(\omega^{II}_{SCE}t)+V_{oc,max}, \nonumber
\end{equation}
\vspace{-0.025in}
where $\omega^{II}_{SCE}=\frac{1}{\sqrt{L_PC_{T,min}}}$ is the resonance frequency of the oscillator. Switch S is kept closed for one-fourth the $L_P$-$C_{T,min}$ oscillation cycle, i.e., for the duration, $T^{II}_{SCE}=\frac{\pi}{2\omega^{II}_{SCE}}$, yielding:
\vspace{-0.025in}
\begin{equation}
V_{C_T}^{II+}:=V_{C_T}(T^{II}_{SCE})=V_{oc,max}. \nonumber
\end{equation}
\vspace{-0.025in}
Note at this point the voltage on the TENG capacitor is the opposite of the TENG voltage source, and the overall
TENG voltage $V_T^{II+}$ drops to zero (by invoking the TENG operating equation (Eq.~(\ref{VT})), while the TENG capacitor charge $Q_{C_T}^{II+}$ can be derived using the capacitor relation:
\vspace{-0.025in}
\begin{align}
    \label{eq:S3}
   &V^{II+}_{T}=V_{oc,max}-V_{C_T}^{II+}=V_{oc,max}-V_{oc,max}=0;\nonumber\\
    & Q_{C_T}^{II+}=C_{T}^{II+}V_{C_T}^{II+}=C_{T,min}V_{oc,max}. 
\end{align}
\vspace{-0.05in}
Thus, during $T^{II}_{SCE}$, the TENG voltage ($V_T(t)$) 
falls from the maxima ($V_{oc,max}$) to zero, while the current in the loop ($I^{II}_{L_P}(t)$) rises sinusoidally from zero to it's maxima ($I^{II}_{L_P,max}$) (refer to green curve Fig.~\ref{SCE_Strip}(c)). During this time period the current in the secondary inductor $L_S$ is blocked by the reverse-biased diode $D$, with the energy transferred from TENG being stored as magnetic energy in the core of the primary inductor $L_p$. On opening the switch S after the quarter of $L_P$-$C_{T,min}$ oscillation time (namely, end of State II+), primary inductor current quickly falls to zero while the secondary inductor current 
shoots up to $I^{II}_{L_P,max}$ (turns ratio = 1, and maintaining the same rate of change). This in turn charges the load battery following the equation $L_S\frac{dI_{L_S}(t)}{dt}=-V_B, t\geq T^{II}_{SCE}$, with initial condition $I_{L_s}(T^{II}_{SCE})=I^{II}_{L_P,max}$. Since the rate of change of current $I_{L_S}(t)$ is constant at $-\frac{V_B}{L_S}$, it decays linearly starting from its peak value $I^{II}_{L_P,max}$ to zero over the time $T^{II}_L:=\frac{I^{II}_{L_P,max}}{V_B/L_S}=\frac{L_SI^{II}_{L_P,max}}{V_B}$. The switching period $T_{SCE}^{II}$ is small compared to the TENG operation period ($T$), and so the TENG capacitance $C_T(t)$ and open circuit voltage $V_{oc}(t)$ are considered unchanged at $C_{T,min}$ and $V_{oc,max}$, respectively, during the switching period.

With the start of second half-cycle, upper plate moves downward with switch S open, to reach the starting position ($x=0$) and State I is achieved, where the two plates are in contact (left bottom of Fig.~\ref{SCE_Strip}(a)). With Switch S open during this downward traversal, the charge on the TENG capacitor is retained from State II+, however the capacitance changes from $C_{T,min}$ to $C_{T,max}$, while the open circuit voltage ($V_{oc}(t)$) falls to zero (Fig.~\ref{StripI}(c)). Thus,
\vspace{-0.025in}
\begin{align}
    \label{eq:S4}
    &Q_{C_T}^{I}=Q_{C_T}^{II+}=C_{T,min}V_{oc,max}; \nonumber\\
    &V^{I}_{T}=V_{oc,min}-V^I_{C_T}=0-\frac{Q_{C_T}^{I}}{C_{T,max}}=-\frac{V_{oc,max}}{\beta}.
    \vspace{-0.05in}
\end{align}
Here, $\beta$ is the ratio of maximum to minimum TENG capacitance as defined earlier. To extract the TENG energy, switch S is closed for one-fourth the $L_P$-$C_{T,max}$ oscillation cycle, entering State I+ (left top of Fig.~\ref{SCE_Strip}(a)), with oscillation frequency and switch S closure time given by,
\begin{equation}
\vspace{-0.05in}
    \label{eq:S5}
    \omega^{I}_{SCE}=\frac{1}{\sqrt{L_PC_{T,max}}}; T^{I}_{SCE}=\frac{\pi}{2\omega^{I}_{SCE}}=\sqrt{\beta}T^{II}_{SCE}.
\end{equation}
With S closed, the SCE circuit is simplified to as shown in the left half of Fig.~\ref{SCE_Strip}(b). Similar to the energy extraction process between State II to State II+, energy is transferred to the primary inductor ($L_P$), followed by to the secondary inductor ($L_S$) (when S is opened $T^I_{SCE}$ time later), and then to the battery load. The full wave rectifier ensures the inductor current ($I^{I}_{L_P}$) polarity is maintained as earlier. At State I+, during the time switch S is closed, the TENG capacitor discharges through the primary inductor (in quarter of $L_P$-$C_{T,max}$ oscillation cycle) and hence $V_{C_T}(t)$ falls to zero. This resets TENG for the start of the next cycle (and State I+ ends):
\begin{equation}
    \label{eq:S6}
    V^{I+}_{T}=V_{oc,min}-V_{C_T}^{I+}=0-0=0;\quad Q_{C_T}^{I+}=0. \nonumber
\end{equation}
\subsection{Per-Cycle Energy Output}\label{SCE_E_Der}
Cyclic TENG Voltage ($V_T(t)$) vs. TENG Capacitor Charge ($Q_{C_T}(t)$) diagram can be used to derive the per-cycle energy output ($E_{cycle}$)\cite{zi2015standards}:

\begin{eqnarray}
\vspace{-0.1in}
    \label{eq:S7}
    E_{cycle}=\int_{0}^{T}V_TI_Tdt=\int_{0}^{T}V_TdQ_{C_T}.
\vspace{-0.1in}
\end{eqnarray}

\begin{figure*}
\vspace*{-.1in}
  \begin{center}
  \includegraphics[width=1.0\linewidth]{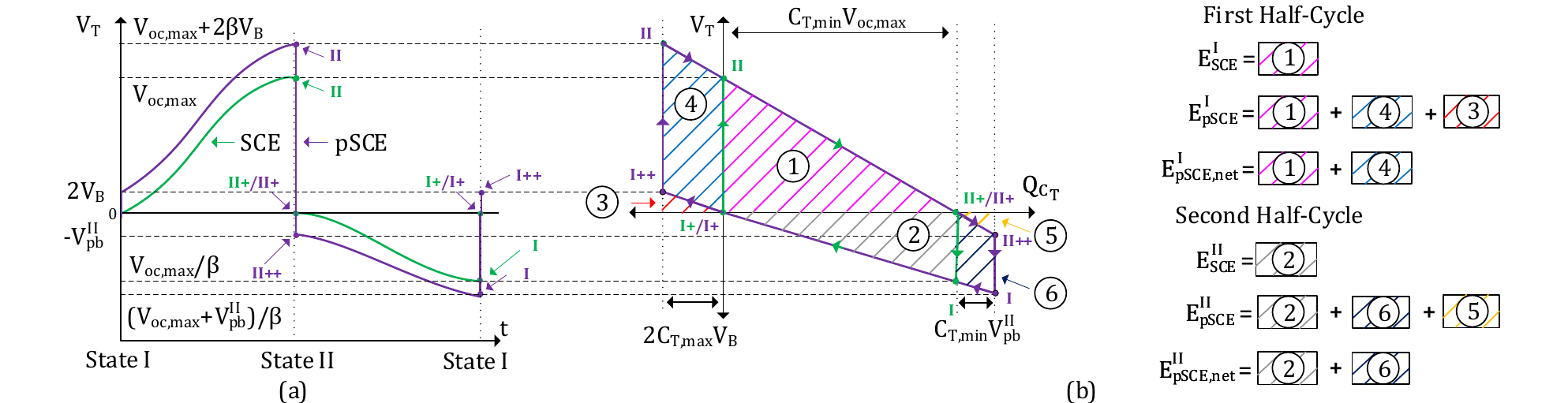}
  \vspace*{-0.2in}
  \caption{(a) TENG voltage waveform plot (green under SCE vs. purple under pSCE), and (e) TENG ``Charge vs. Voltage" plot for SCE (areas 1-2) and pSCE (areas 1-6).}\label{V_Q_Plot}
  \end{center}
\vspace*{-.35in}
\end{figure*}

Above equation shows that the energy output during the extraction step is equal to the area enclosed by the $V_T$ curve against the $Q_{C_T}$ axis. Using, the TENG voltage ($V_T$) and TENG capacitor charge ($Q_{C_T}$) derived at State I, I+, II and II+ in the above section, the $V_T$ vs. $Q_{C_T}$ plot for SCE is obtained in Fig.~\ref{V_Q_Plot}(b) as explained next. At State I$+$ both $V_T$ and $Q_{C_T}$ are reset to zero. Then in moving from I$+$ to II (separation of plates from minimum to maximum), $Q_{C_T}$ stays at zero (no change in capacitor charge), while $V_T$ rises (vertically since $Q_{C_T}$ is held constant at zero) to maximum ($=V_{oc,max}$). Then in moving from II to II$+$, $V_T$ drops to zero (as $V_{C_T}$ rises sinusoidally to $V_{oc,max}$), while $Q_{C_T}$ becomes $C_{T,min}V_{oc,max}$; the corresponding $V_T$ vs. $Q_{C_T}$ change is linear since during this period, $\frac{dV_T}{dQ_{C_T}}=\frac{d(V_{oc,max}-V_{C_T})}{dQ_{C_T}}=-\frac{dV_{C_T}}{dQ_{C_T}}=-\frac{1}{C_{T,min}}$. The move from II$+$ to I (plates moving from maximally separated to being in contact) is the reverse of that from I$+$ to II, where $Q_{C_T}$ is again held constant at $C_{T,min}V_{oc,max}$, while the $V_{oc}$ part of $V_T$ drops to zero and the $V_{C_T}$ part of $V_T$ goes from $V_{oc,max}$ to $V_{oc,max}/\beta$ (since $Q_{C_T}$ is conserved  at $C_{T,min}V_{oc,max}$ but the capacitance changes from $C_{T,min}$ to $C_{T,max}$). Finally, the behavior from State I to I$+$ to similar to that of State II to II$+$, except the slope is $-\frac{1}{C_{T,max}}$ (as opposed to $-\frac{1}{C_{T,min}}$ since the capacitance during this time is $C_{T,max}$ as opposed to $C_{T,min}$).
Then there are two energy extraction steps per-cycle: at the end of the first half-cycle, II-II+, extracting energy $E_{SCE}^{I}$ (area of region `1' in Fig.~\ref{V_Q_Plot}(b)), and at the end of the second half-cycle, I-I+, extracting energy $E_{SCE}^{II}$  (area of region `2' in Fig.~\ref{V_Q_Plot}(b)). Thus, the energy output can be found by computing the enclosed triangular areas of the regions `1' and `2' in Fig.~\ref{V_Q_Plot}(b) as:
\vspace{-0.05in}
\begin{align}
\vspace{-0.05in}
    \label{eq:S8}
    &E_{SCE}^{I}=\frac{1}{2}\times{C_{T,min}V_{oc,max}}\times{V_{oc,max}}; \nonumber\\
    &E_{SCE}^{II}=\frac{1}{2}\times{C_{T,min}V_{oc,max}}\times{\frac{V_{oc,max}}{\beta}}; \nonumber\\
    &E_{SCE}=E_{SCE}^{I}+E_{SCE}^{II}=\frac{1}{2}\left(\!1+\frac{1}{\beta}\!\right)\!C_{T,min}V^2_{oc,max}.
\end{align}
The $V_T$ vs. $Q_{C_T}$ plot for the one SCE cycle matches that of CMEO \cite{zi2015standards,cheng2017high}. \begin{r1}\label{Passive_EEC}Thus, all {\em passive} EECs (i.e., barring the {\em active} ones) are limited to the area enclosed by the SCE's $V_T$ vs. $Q_{C_T}$ trapezoid, and in this sense, SCE is optimal and also maximally efficient among all the passive EECs.
Also, SCE's energy output (Eq.~(\ref{eq:S8})) is independent of the load battery value, which is another nice feature of SCE. We also note that in practical implementation of the SCE circuit, the energy delivered to the load is a fraction of the ideal extracted energy (Eq.~(\ref{eq:S8})) due to different circuit parasitic losses. Hence, we have provided the derivation of the delivered per-cycle energy considering the resistive loss of the non-ideal inductor in Supplementary Note~S.I available at \cite{pathak2021prebias}\cite{pathak2021prebias_arxiv}.
\end{r1}
\vspace{-0.1in}
\section{Active Pre-biasing for Enhanced Extraction}\label{Pre-bias_Sec}
Pre-biasing refers to charging the TENG capacitor before the start of each half-cycle to be able to possibly increase the energy output beyond the SCE output (CMEO). {\em Active} pre-biasing uses previously extracted energy for this purpose. Passive pre-biasing is also possible as in the case of P-SSHI and S-SSHI EECs \cite{pathak2018modeling,xu2019boost,pathak2021synchronous}, but those do not fully extract the transduced energy. The active pre-biasing approach we present below retains the SCE feature that it extracts the full amount of transduced energy, and show that it is the most efficient (among the known classes of EECs).

\subsubsection{First half-cycle} 
\begin{figure*}[h]
\vspace*{-.1in}
  \begin{center}
  \includegraphics[width=0.95\linewidth]{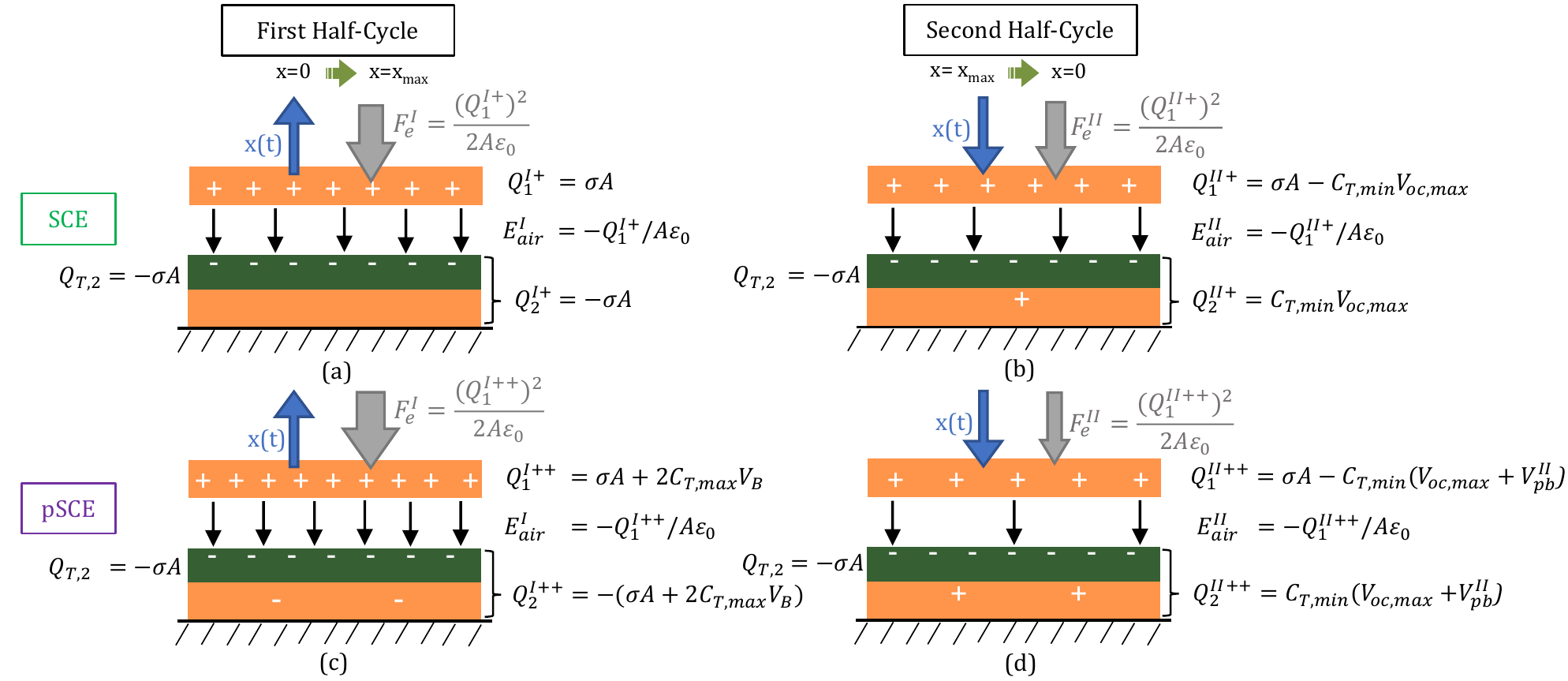}
  \vspace{-0.15in}
  \caption{Schematic of TENG mechanical motion ($x(t)$) and electrostatic force ($F_e$) for the SCE circuit during the (a) first half-cycle and (b) second half-cycle and, for the pSCE circuit during the (c) first half-cycle and (d) second half-cycle.}\label{Force}
  \end{center}
\vspace*{-.325in}
\end{figure*}

For the SCE circuit, consider the first half-cycle, i.e., starting at State I+ ($x=0$). Then as shown in the above section, the charge on the TENG capacitor ($Q_{C_T}^{I+}=0$), and hence the charge on the upper plate ($Q_1^{I+}$) and the lower plate ($Q_2^{I+}$) are given by,
\vspace{-0.05in}
\begin{equation}
    \label{eq:E1}
    Q_1^{I+}=Q_{T,1}-Q_{C_T}^{I+}=\sigma A; \quad Q_2^{I+}=-Q_1^{I+}=-\sigma A.
\end{equation}
\vspace{-0.05in}
In transitioning from State I+ ($x=0$) to State II ($x=x_{max}$), external mechanical excitation moves up the upper plate against the electrostatic force of attraction ($F_e^{I}$) acting on it (Refer Fig.~\ref{Force}(a)). Noting that the electric field due to the bottom plate that acts on the charges of the upper plate is $\frac{ 1}{2}\vec E_\sigma$, using Eq.\ref{E'}, we get:
\begin{align}
    \label{eq:E2}
    \Vec{F_e^{I}}&=\frac{1}{2}\vec E_\sigma Q_1^{I+}=-\frac{1}{2}\left(\frac{\sigma}{\epsilon_0}\vec 1_x\right)(\sigma A)=-\frac{1}{2}\left(\frac{\sigma^2 A}{\epsilon_0}\right)\vec 1_x;\nonumber\\
    W_e^{I}&=\int_{0}^{x_{max}}\!\!{\vec F_e^{I}}. d\vec \!x=\!-\frac{1}{2}\!\left(\!\!\frac{\sigma^2 A}{\epsilon_0}\!\!\right)\!\!x_{max}\!=\!-\frac{1}{2}\!\left(\!\!\frac{(Q_1^{I+})^2 }{\epsilon_0A}\!\!\right)\!\!x_{max}\nonumber \\
   &=-\frac{1}{2}\!\left(\frac{\beta}{\beta-1}\right)\!C_{T,min}V_{oc,max}^2,
\end{align}
where in the last equality we use the facts that $\left(\!\!\frac{\beta}{\beta-1}\!\!\right)\!C_{T,min}=\frac{C_{T,max}}
{\frac{C_{T,max}}{C_{T,min}}-1}=
\frac{\epsilon_0A}{x_{max}}$ (using $C_{T,min}=\frac{\epsilon_0 A}{x_{max}+d_{eff}};C_{T,max}=\frac{\epsilon_0 A}{d_{eff}}$), and $V_{oc,max}=\frac{\sigma x_{max}}{\epsilon_0}$ so that $\left(\!\!\frac{\beta}{\beta-1}\!\!\right)\!C_{T,min}V^2_{oc,max}=\frac{\epsilon_0A}{x_{max}}\!\left(\!\!\frac{\sigma x_{max}}{\epsilon_0}\!\!\right)^2\!=\!\left(\!\!\frac{\sigma^2 A}{\epsilon_0}\!\!\right)\!x_{max}$.
The work $W_e^I$ done by the mechanical source (stored as electrical potential energy) is proportional to the electrostatic force which in turn is proportional to the square of the charge on the upper electrode ($(Q_1^{I+})^2$). Thus, by pre-biasing (i.e., adding extra charge) the TENG at State I+, $Q_1^{I+}$ can be increased to in-turn increase the transduced energy. This establishes the motivation behind pre-biasing, and in the following section we quantitatively derive the increased transduced energy for the proposed pSCE architecture. 

\begin{r1}\label{pSCE_EI}
It is clear that the transduced energy in the first half-cycle will continue to increase on increasing the pre-biased charge.
In practice, the upper limit may be set by the available external mechanical force (equivalently acceleration) that is countered by electrostatic force of attraction. For a given periodic mechanical force with fixed acceleration, increasing the level of pre-biasing will increase the countering electrostatic force that will eventually reduce the maximum reachable separation between the two plates ($x_{max}$) and, as a result, decrease the maximum open circuit TENG source voltage ($V_{oc,max}$). This effect of feedback (electrostatic force) in a vibration driven electrostatic transducer on its mechanical motion and the transduced energy has been previously studied in works such as  \cite{galayko2010general,mitcheson2012maximum,karami2015limiting}. In a typical TENG application with a moving part (upper plate in our case) of mass $m$, 
the deceleration due to electrical attraction ($F_e/m$) is an order or two lower in magnitude than the external mechanical force (equivalently acceleration). For example, human walking provides 2-3~$m/s^2$ acceleration\cite{li2014energy}, while a car engine compartment provides $12\ m/s^2$\cite{li2014energy}. In contrast, for the TENG used in this work (moving mass, $m=113.7g$), the electrical deceleration is $0.013\ m/s^2$ with no pre-biasing (SCE operation) and $0.026\ m/s^2$ when pre-biased at 30~V as shown in Supplementary Note~S.IV\cite{pathak2021prebias}\cite{pathak2021prebias_arxiv}. Thus, pre-biasing does not outpower the mechanical acceleration to cause a appreciable reduction in $x_{max}$ value. Thereby in most meso-scale implementations, pre-biasing level is limited only by the upper limit of the voltage source available for pre-biasing, air electric field breakdown or the voltage ratings of the components like the switches used in the pSCE circuit, as discussed in Sec.~\ref{Results}.
\end{r1}

\subsubsection{Second half-cycle} At the end of the first half-cycle, part of the transduced energy is extracted and delivered to the load ($E_{SCE}^{I}$), while part of it remains stored on the TENG capacitor. The charge on the upper plate at this stage, i.e., at State II+ ($Q_1^{II+}$) can be found using Eq.~(\ref{eq:S3}):
\begin{equation}
    \label{eq:E3}
    Q_1^{II+}=Q_{T,1}-Q_{C_T}^{II+}=\sigma A-C_{T,min}V_{oc,max}.
\end{equation}
With the beginning of second half-cycle, as shown in Fig.~\ref{Force}(b), direction of upper plate motion is reversed to match that of electrostatic force. This results in positive work or in other words a part of stored TENG capacitor energy is dissipated back into the environment. This energy loss can be derived as in Eq.~(\ref{eq:E2}):
\begin{align}
    \label{eq:E4}
    W_e^{II}\!&\!=\!\frac{1}{2}\!\left(\!\!\frac{(Q_1^{II+})^2 }{\epsilon_0A}\!\!\right)\!\!x_{max}\!=\!\frac{1}{2}\!\left(\!\!\frac{(\sigma A-C_{T,min}V_{oc,max})^2 }{\epsilon_0A}\!\!\right)\!\!x_{max}\nonumber \\
   &\!=\!\frac{1}{2}\!\left(\!\frac{1}{\beta(\beta-1)}\!\right)\!C_{T,min}V_{oc,max}^2,
\end{align}
where in deriving the last equality again we have used $C_{T,min}=\frac{\epsilon_0 A}{x_{max}+d_{eff}};C_{T,max}=\frac{\epsilon_0 A}{d_{eff}};V_{oc,max}=\frac{\sigma x_{max}}{\epsilon_0}$.

In the entire cycle, the net work done (the work done in the first half cycle minus the energy lost in the second half cycle) is extracted. Indeed adding the above energy loss $W_e^{II}$ (Eq.~(\ref{eq:E4})) to the extracted energy $E_{SCE}$ (Eq.~(\ref{eq:S8})) gives us the transduced energy of the first cycle $W_e^I$ (Eq.~(\ref{eq:E2})), since the three multipliers of the common term $\frac{1}{2}C_{T,min}V_{oc,max}^2$ appearing in $W_e^{II},E_{SCE},W_e^{I}$ satisfy: \[\left(\!\!\frac{1}{\beta(\beta-1)}\!\!\right)+\left(\!\!1+\frac{1}{\beta}\!\!\right)=\left(\!\!\frac{\beta}{\beta-1}\!\!\right)
\Rightarrow |W_e^{II}+E_{SCE}|=|W_e^I|,\] verifying the energy balance. We make a note that this the first time such an energy balance among the mechanical energy transduced, mechanical energy lost, and electrical energy extracted for a TENG with SCE as EEC has been demonstrated to the best of knowledge.

For reducing the above energy loss to zero (minimum possible), pre-bias charge should be added to turn the plate charges ($Q_1^{II+}=-Q_2^{II+}$) to zero. 
Using Eq.~(\ref{eq:E3}),
\vspace{-.025in}
\begin{align}
    \label{eq:E5}
    Q_1^{II++}=0 &\Rightarrow \sigma A-C_{T,min}V_{oc,max}-Q_{pb,opt}^{II}=0 \nonumber\\
    &\Rightarrow V_{pb,opt}^{II}:=\frac{Q_{pb,opt}^{II}}{C_{T,min}}=\frac{\sigma A}{C_{T,min}}-V_{oc,max} \nonumber\\
    &\Rightarrow V_{pb,opt}^{II}=\frac{V_{oc,max}}{\beta-1},
\end{align}
\vspace{-.025in}
\begin{figure*}
\vspace*{-.1in}
  \begin{center}
  \includegraphics[width=0.95\linewidth]{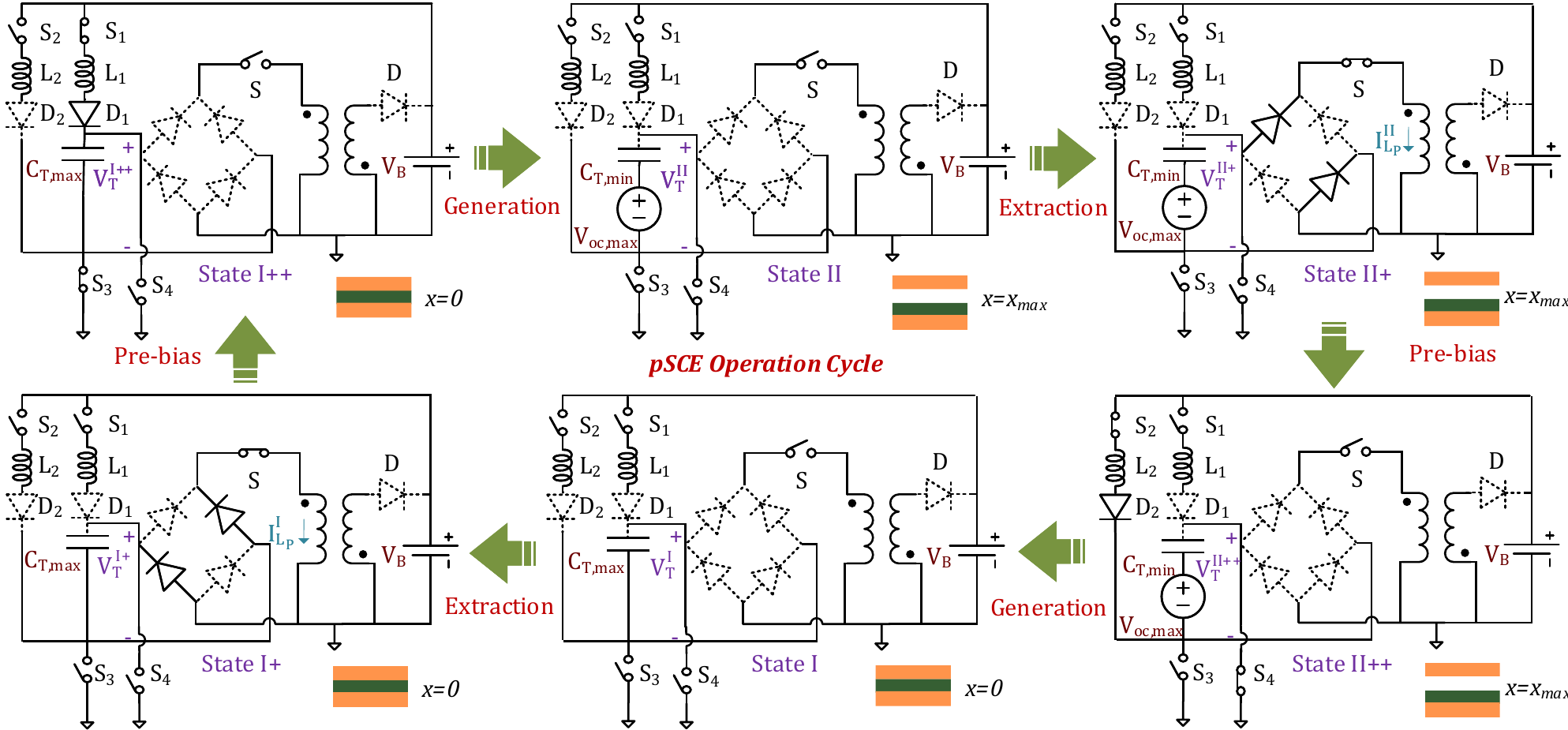}
  \vspace*{-.10in}
  \caption{pSCE circuit operation cycle}\label{pSCE_Operation}
  \end{center}
\vspace*{-.25in}
\end{figure*}
where the last equality follows from deriving the equality,
\vspace*{-.025in}
\begin{equation}\label{eq:relation1}
    \frac{\sigma A}{C_{T,min}}=\frac{\beta}{\beta -1}V_{oc,max},
\end{equation} 
\vspace*{-.025in}
as follows: Since $\beta=\frac{x_{max}+d_{eff}}{d_{eff}}$, we have $\frac{\beta}{\beta-1}=\frac{x_{max}+d_{eff}}{x_{max}}=\frac{(\epsilon_0 A)/C_{T,min}}{(\epsilon_0 V_{oc,max})/\sigma}=\frac{\sigma A}{C_{T,min} V_{oc,max}}$.
It should be noted that the pre-bias voltage beyond twice the above derived optimum value ($V_{pb,opt}^{II}$) would decrease $Q_1^{II++}$ below $-[\sigma A-C_{T,min}V_{oc,max}]$ (and contrastly would increase $Q_2^{II++}=-Q_1^{II++}$ above $[\sigma A-C_{T,min}V_{oc,max}]$), increasing the electrostatic attraction between the two plates beyond SCE (recall, the charge on the two plates at State II+ in case of SCE is $\pm[\sigma A -C_{T,min}V_{oc,max}]$), thereby causing a higher loss to environment in the second half cycle than that of SCE (as given by $W_e^{II}=\frac{1}{2}\!\left(\!\!\frac{(Q_1^{II+})^2 }{\epsilon_0A}\!\!\right)\!\!x_{max}$ (Eq.~(\ref{eq:E4})). Thus, 
\begin{equation}\label{eq:pbu}
V^{II}_{pb,u}:=2V^{11}_{pb,opt}=\frac{2V_{oc,max}}{\beta-1}
\end{equation}
provides the upper limit to pre-biasing in the second half cycle for it to remain favorable over SCE. In contrast, pre-biasing is always favorable in the first half cycle and also overall, combining the two cycles, as demonstrated in the next section. 
\vspace{-0.2in}
\section{Proposed pSCE and its Analysis}\label{Sec_pSCE}
\vspace{-0.05in}
Here we describe the proposed circuit that is used for pre-biasing the TENG at the start of each half-cycle. To enable the same with either polarity, four additional switches S1-S4 configured as H-bridge are added to the SCE circuit architecture (Fig.~\ref{Circuit_Diagram}).
\begin{figure}[htbp]
\vspace*{-.05in}
  \begin{center}
  \includegraphics[width=1.0\linewidth]{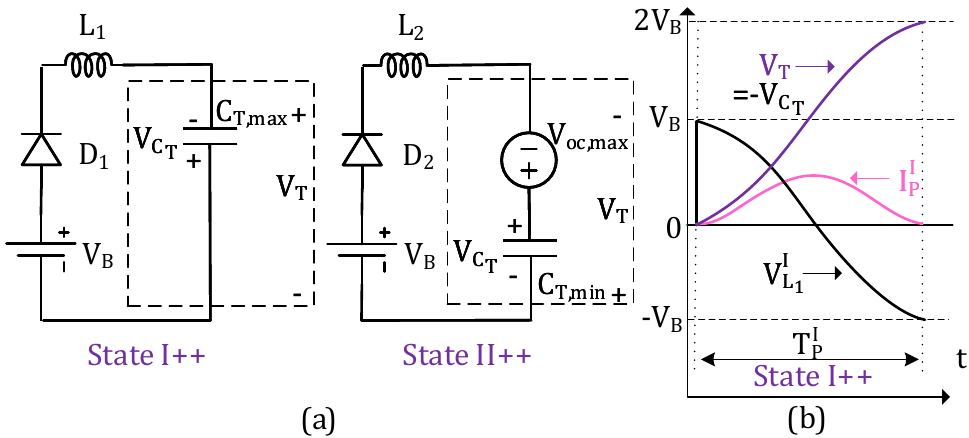}
  \vspace{-.225in}
  \caption{pSCE circuit (a) Simplified circuit during pre-biasing at start of both the half-cycle and (b) $L_1$ Inductor Voltage ($V_{L_1}$), Loop Current ($I_P$) and, TENG Voltage ($V_T$) during pre-biasing at State I++.}\label{Prebias_Simplified}
  \end{center}
\vspace*{-.25in}
\end{figure}
\vspace*{-0.2in}
\subsection{pSCE operation} \label{pSCE_operation_description}
pSCE operation is obtained by extending the SCE operation discussed above with two added states for pre-biasing at start of each half-cycle: State I++ (following I+) and State II++ (following II+). Circuit operation can be understood by following the operation cycle diagram of Fig.~\ref{pSCE_Operation} and TENG voltage waveform of Fig.~\ref{V_Q_Plot}(a). As in SCE, the operation commences at $x=0$, with all switches open and $V_{oc}=0$ (State I+). Then for pre-biasing, switches S1 and S3 are closed (State I++ at the left top in Fig.~\ref{pSCE_Operation}) to form the pre-bias charging loop, $V_B-S_1-L_1-D_1-C_T-S_3$ with the circuit simplified to a $L_1$-$C_{T,max}$ resonator loop as shown in left half of Fig.~\ref{Prebias_Simplified}(a). The corresponding circuit differential equation is similar to Eq.~(\ref{eq:D1}). The TENG capacitor voltage $V_{C_T}(t)$ with initial condition $V_{C_T}(0)=I_{C_T}(0)=0$ is obtained as,
\vspace*{-.025in}
\begin{equation}
V_{C_T}(t)=-V_{B}(cos(\omega^{I}_{P}t))+V_{B}; \quad \omega^{I}_{P}=\frac{1}{\sqrt{L_1C_{T,max}}}. \nonumber
\end{equation}
 \vspace*{-.025in}
The switches S1 and S3 are kept closed for half the resonance oscillation cycle ($T_P^{I}=\frac{\pi}{\omega^{I}_P}$) to reach  the TENG capacitor voltage of $2V_B$:
\begin{equation}
V_{C_T}^{I++}:=V_{C_T}(T_P^{I})=2V_{B}. \nonumber
\end{equation}
Using the TENG operating equation (Eq.~(\ref{VT})), with $V_{oc}=0$,
\vspace*{-.025in}
\begin{equation}
    \label{eq:T1}
    V^{I++}_{T}=-\frac{Q_{C_T}^{I++}}{C_{T,max}}=2V_B
    \Rightarrow Q_{C_T}^{I++}=-2C_{T,max}V_B. 
\end{equation}
\vspace*{-.025in}
Thus, using the $L_1$-$C_T$ resonance, the pre-bias voltage is passively boosted to twice the battery voltage acting as pre-biasing source, by introducing the extra inductor $L_1$ (correspondingly $L_2$ for second half-cycle). At this point, the two plates separate with switch S open, and TENG voltage increases to reach it's maxima at State II. Since, no current has flown through TENG since the last state, $Q_{C_T}^{II}=Q_{C_T}^{I++}$. Hence,
\vspace*{-.025in}
\begin{equation}
    \label{eq:T2}
     V^{II}_{T}=V_{oc,max}-\frac{Q_{C_T}^{II}}{C_{T,min}}=V_{oc,max}+2\beta V_B.
\end{equation}
\vspace*{-.025in}
Note the increase in the voltage by $2\beta V_B$ compared to the SCE circuit. At this stage, similar to the SCE circuit, switch S is closed for one-fourth the $L_P$-$C_{T,min}$ oscillation cycle ($T_{SCE}^{II}$). TENG voltage falls to zero (State II+) and the energy is transferred to the primary side inductor $L_p$ which is subsequently transferred to load battery via secondary inductor $L_S$. With the first half-cycle complete, similar to the State I+ to I++ operation, the State II+ to II++ operation is performed to pre-bias the TENG by closing S2 and S4 (State II++ at the right bottom in Fig.~\ref{pSCE_Operation}) to form the $V_B-S_2-L_2-D_2-C_T-S_4$ loop, with the circuit simplified to a $L_2$-$C_{T,min}$ resonator as shown in the right half of Fig.~\ref{Prebias_Simplified}(a), with initial conditions: $V_{C_T}(0)=V_{oc,max};I_{C_T}(0)=0$, and the TENG capacitor satisfying: 
\vspace*{-.025in}
\begin{equation}
V_{C_T}(t)=V_{B}(1-cos(\omega^{II}_{P}t))+V_{oc,max}; \- \omega^{II}_{P}=\frac{1}{\sqrt{L_2C_{T,min}}}. \nonumber
\end{equation}
\vspace*{-.025in}
Note that $V_{C_T}(t)$ oscillates between $V_{oc,max}$ and $V_{oc,max}+2V_B$, and so the TENG voltage $V_T(t)=V_{oc,max}-V_{C_T}(t)$ oscillates between 0 and $-2V_B$. Thus by controlling the duration over which S2 and S4 are closed, i.e., the duration of pre-biasing, the TENG voltage can be set to any value $-V^{II}_{pb}$ such that $-2V_B\leq -V^{II}_{pb}\leq 0$ (Refer Supplementary Note~S.V at \cite{pathak2021prebias}\cite{pathak2021prebias_arxiv} for further details). Then,
\vspace*{-.025in}
\begin{align}
    \label{eq:T3}
    &V^{II++}_{T}=V_{oc,max}-\frac{Q_{C_T}^{II++}}{C_{T,min}}=-V^{II}_{pb} \nonumber\\
    \Rightarrow &Q_{C_T}^{II++}=C_{T,min}(V_{oc,max}+V^{II}_{pb}).
\end{align}
\vspace*{-.025in}
Next, as per the periodic motion, the plates come together ($x=0$) as in State I. Since, the TENG is in open circuit condition during this movement, $Q_{C_T}^{I}=Q_{C_T}^{II++}$. Hence,
\vspace*{-.025in}
\begin{equation}
    \label{eq:T4}
     V^{I}_{T}=-\frac{Q_{C_T}^{I}}{C_{T,max}}=-\frac{V_{oc,max}+V^{II}_{pb}}{\beta}.
\end{equation}
\vspace*{-.025in}
Again, at this stage, the TENG voltage is higher (in absolute terms) over the SCE circuit (see Eq.~(\ref{eq:S4})), meaning a larger energy will be recovered in going from State I to State I+. Next, the switch S is closed for one-fourth the $L_p$-$C_{T,max}$ oscillation cycle ($T_{SCE}^{I}$) to extract the energy from TENG (State I+ at the left bottom in Fig.~\ref{pSCE_Operation}). TENG voltage falls to zero completing one full operation cycle.
\vspace*{-0.15in}
\subsection{Per-Cycle Energy Output}
As with the case of SCE per-cycle energy output calculation, we plot the TENG voltage ($V_T$) and TENG capacitor charge ($Q_{C_T}$), derived above, at different states to obtain the $V_T$-$Q_{C_T}$ plot of Fig.~\ref{V_Q_Plot}(b). Note due to pre-biasing, the trapezoidal area of SCE extends on both sides (beyond the regions labeled`1' and `2' of SCE), adding the regions labeled as `3', `4', `5', and `6'. The pre-biasing adds charge $-2C_{T,max}V_B$ raising TENG voltage to $2V_B$ at State I++ (Eq.~(\ref{eq:T1})), while adds charge $C_{T,min}V^{II}_{pb}$ lowering TENG voltage to $-V^{II}_{pb}$ at State II++ (Eq.~(\ref{eq:T3})). These are also marked in the $V_T$-$Q_{C_T}$ plot of Fig.~\ref{V_Q_Plot}(b). As a result, the energy extracted at the end of the first half-cycle ($E^I_{pSCE}$) is equal to the triangular area enclosed by the $V_T$-$Q_{C_T}$ curve of the extraction step, namely, State II to II+ line, and the $Q_{C_T}$ axis (areas `1'+`3'+`4' in Fig.~\ref{V_Q_Plot}(b)). Similarly, energy extracted at the end of the second half cycle ($E^{II}_{pSCE}$) is the triangular area between State I to I+ line and the $Q_{C_T}$ axis (areas `2'+`5'+`6' in Fig.~\ref{V_Q_Plot}(b)):
\begin{align}
    \label{eq:T5}
    &E_{pSCE}^{I}\!=\!\frac{1}{2}\!\times\!{C_{T,min}(V_{oc,max}}\!+\!2\beta V_B)\!\times\!({V_{oc,max}}\!+\!2\beta V_B); \nonumber\\
   &E_{pSCE}^{II}\!=\!\frac{1}{2}\!\times\!{C_{T,min}(V_{oc,max}\!+\!V^{II}_{pb})}\!\!\times\!\!{\frac{(V_{oc,max}\!+\!V^{II}_{pb}\!)\!}{\beta}}.
\end{align}
The energy consumed from the load battery for the TENG pre-bias at the start of each half-cycle is equal to the product of the battery voltage and the charge flow during the pre-biasing periods. 
The consumed pre-biasing energy for the first half-cycle ($E_{pre{\mbox -}bias}^I$) and that for the second half-cycle ($E_{pre{\mbox -}bias}^{II}$) are thus respectively equal to the areas marked as `3’ and `5’ on the $V_T$-$Q_{C_T}$ plot of Fig.~\ref{V_Q_Plot}(b):
\vspace*{-.025in}
\begin{align}
    \label{eq:T6}
    &E^{I}_{pre{\mbox -}bias}=2C_{T,max}V_B^2=2\beta C_{T,min}V_B^2; \nonumber\\ 
    &E^{II}_{pre{\mbox -}bias}=\frac{1}{2}C_{T,min}(V^{II}_{pb})^2.
\end{align}
\vspace*{-.025in}
Thus, the net energy delivered to the battery load in the two half-cycles  $E_{pSCE,net}^{I}=E_{pSCE}^{I}-E_{pre{\mbox -}bias}^{I}$, $E_{pSCE,net}^{II}=E_{pSCE}^{II}-E_{pre{\mbox -}bias}^{II}$, and over the entire cycle $E_{pSCE,net}=E_{pSCE,net}^{I}+E_{pSCE,net}^{II}$ are given by:
\begin{align}
    \label{eq:T7}
    E_{pSCE,net}^{I}=&\frac{1}{2}C_{T,min}[(V_{oc,max}+2\beta V_B)^2-4\beta V_B^2]; \nonumber\\
  E_{pSCE,net}^{II}=&\frac{1}{2}C_{T,min}[\frac{(V_{oc,max}+V^{II}_{pb})^2}{\beta}- (V^{II}_{pb})^2];\nonumber \\
   E_{pSCE,net}=&\frac{1}{2}C_{T,min}[(V_{oc,max}+2\beta V_B)^2\nonumber\\
   &+\!\frac{(V_{oc,max}+V^{II}_{pb})^2}{\beta}\!-\!4\beta V_B^2\!-\!(V^{II}_{pb})^2].
\vspace*{-0.15in}
\end{align}
Above is equal to the area enclosed by the extended trapezoid (regions `4'+`1'+`2'+`6') of the pSCE operation $V_T$-$Q_{C_T}$ plot shown in Fig.~\ref{V_Q_Plot}(b) and represents the per-cycle energy extracted from TENG. However, the energy delivered to the load will be lowered due to inductor resistive loss in the extraction and pre-biasing (H-bridge) circuit marked as “Loss 2” and “Loss 3” in Fig.~\ref{Energy_Flow}, respectively. For interested readers, the derivation of the delivered energy with consideration of these losses is provided in the Supplementary Note~S.II of this work's extended version \cite{pathak2021prebias}\cite{pathak2021prebias_arxiv}. 
\vspace*{-0.15in}
\subsection{Conditions for Pre-biasing}\label{condition}
\vspace*{-0.05in}
For the first half-cycle, {\em Remark~\ref{pSCE_EI}} above discussed that compared to the SCE operation, pre-biasing increases the transduced energy which can now be verified from energy point of view by showing that  $E_{pSCE,net}^{I}$ at any arbitrary pre-biasing voltage (substituting $V_{pb}^{I}$ for $2V_B$ in Eq.~(\ref{eq:T7})) is greater than $E_{SCE}^{I}$ (Eq.~(\ref{eq:S8})) and continues to increase with $V_{pb}^{I}$:
\begin{align*}
\vspace*{-0.25in}
    &E_{pSCE,net}^{I}-E_{SCE}^{I}\geq 0\nonumber \\
    \Leftrightarrow&\left[(V_{oc,max}+\beta V^{I}_{pb})^2-\beta(V^{I}_{pb})^2\right]-V_{oc,max}^2\geq 0 \nonumber\\
    \Leftrightarrow&\beta\left[2V_{oc,max}+(\beta-1)V^{I}_{pb}\right]V^{I}_{pb}\geq 0 \nonumber
\vspace*{-0.15in}
\end{align*}
Since $\beta\geq 1$, the above inequality clearly holds, where the left hand side is an increasing function of $V_{pb}^{I}$ (higher the pre-biasing; higher the gain).  

For the second half-cycle, we noted in Sec.~\ref{Pre-bias_Sec} that pre-biasing beyond a limit increases the energy loss to the environment so it becomes higher than the corresponding energy loss of SCE. Here we show that the same limit on the battery load, beyond which it is preferable to skip pre-biasing for the second half-cycle, can be derived by the way of energy considerations:
\begin{align} \label{eq:C2}
\vspace*{-0.05in}
    &E_{pSCE,net}^{II}-E_{SCE}^{II}\geq 0\nonumber \\
    \Leftrightarrow&\left[\frac{(V_{oc,max}+V^{II}_{pb})^2}{\beta}-(V^{II}_{pb})^2\right]-\frac{V_{oc,max}^2}{\beta}\geq 0 \nonumber\\
    \Leftrightarrow&2V_{oc,max}-(\beta-1) V^{II}_{pb}\geq 0 \nonumber\\
        \Leftrightarrow&V^{II}_{pb}\leq \frac{2V_{oc,max}}{(\beta-1)}=:V^{II}_{pb,u}.
\vspace*{-0.05in}
\end{align}
Note this condition obtained based on energy considerations matches the condition of Eq.~(\ref{eq:pbu}) that was derived considering the level of charge on the two plates (and the corresponding level of electrostatic attraction), providing a correctness check to our derivations.

The condition of Eq.~(\ref{eq:C2}) can also be visualized on the $V_T$-$Q_{C_T}$ plot as the area of the region marked ‘6’ (net energy gain) being greater than the area of ‘5’ (invested energy). Since the two sloping lines (of slopes $-\frac{1}{C_{T,min}}$ vs. $-\frac{1}{C_{T,min}}$) forming these two regions are convergent in the fourth quadrant on extension, there's an upper limit on pre-biasing when two areas turn equal to each other, which is the condition of Eq.~(\ref{eq:C2}) derived above. 

It turns out that, smaller than the above upper limit, there exists an optimum value of pre-biasing when the difference between the two areas is the largest. Note since $E^{II}_{SCE}=\frac{(V_{oc,max})^2}{\beta}$ is independent of pre-biasing level $V_{pb}$, the optimum of the energy difference $E_{pSCE,net}^{II}-E_{SCE}^{II}$ is obtained by maximizing $E_{pSCE,net}^{II}=\frac{(V_{oc,max}+V^{II}_{pb})^2}{\beta}-(V^{II}_{pb})^2$ part with respect to $V^{II}_{pb}$. Differentiating $E_{pSCE,net}^{II}$ with respect to $V^{II}_{pb}$ yields: $\frac{2(V_{oc,max}+V^{II}_{pb})}{\beta}-2(V^{II}_{pb})$, and equating it to zero provides the optimum level of pre-biasing:
\vspace*{-.025in}
\begin{equation}
    \label{eq:C3}
    V^{II}_{pb,opt}=\frac{V_{oc,max}}{(\beta-1)}.
\end{equation}
\vspace*{-.025in}
The above value, derived from energy considerations, is consistent with the result of Eq.~(\ref{eq:E5}) derived from the charges/forces perspective in Sec.~\ref{Pre-bias_Sec}, affirming again the correctness of our derivations. Applying Eq.~(\ref{eq:T3}) to the case of optimized pre-biasing (i.e., substituting $V^{II}_{pb,opt}$ for $V^{II}_{pb}$ in Eq.~(\ref{eq:T3})), we get the TENG capacitor charge to be, $Q^{II++}_{C_T}=C_{T,min}[V_{oc,max}+\frac{V_{oc,max}}{\beta-1}]=(\frac{\beta}{\beta-1})C_{T,min}V_{oc,max}$. 
Then the total charge on the upper plate will be given by subtracting the above capacitive charge from the triboelectric charge $\sigma A$, namely, \[Q_1^{II++}=\sigma A-(\frac{\beta}{\beta-1})C_{T,min}V_{oc,max}=0,\]
where the last equality follows from Eq.~(\ref{eq:relation1}). This is indeed another pleasing result confirming our earlier insight that setting the TENG charge to zero at State II++ is the best possible pre-biasing for minimizing the loss $W^{II}_e$ (see Eq.~(\ref{eq:E5})). 

\begin{r1}\label{pSCE_Opt}Since $V^{II}_{pb,opt}<V^{II}_{pb,u}$ (the former is the half of the latter), it follows that it is favorable to operate at the optimum pre-bias level of $V^{II}_{pb,opt}$ in the second half cycle whenever it is feasible. So if  $V^{II}_{pb,opt}\leq 2V_B$, then in the second half cycle, one would pre-bias the TENG to the voltage $-V^{II}_{pb,opt}$ by appropriately controlling the on-time of the switches S2 and S4 (note as discussed just prior to Eq.~(\ref{eq:T3}), it is possible to pre-bias the TENG voltage to as low as $-2V_B$). On the other hand, if $2V_B< V^{II}_{pb,opt}$, then the best one can do from the energy perspective in the second half cycle is to pre-bias the TENG to the voltage $-2V_B$.

Another novel result can be deduced by mapping the derived optimal battery voltage value $V^{II}_{pb,opt}$ of Eq.~(\ref{eq:C3}) to the TENG parameters: \[\frac{V_{oc,max}}{\beta-1}=\frac{\frac{\sigma x_{max}}{\epsilon_0}}{\frac{x_{max}+d_{eff}}{d_{eff}}-1}=\frac{\sigma d_{eff}}{\epsilon_0},\] 
which is a constant value for a given TENG. This implies that the optimal battery load $V^{II}_{pb,opt}$ value doesn't depend on the operation parameters such as amplitude ($x_{max}$) or frequency of operation, and as such no dynamic closed-loop control is required to adjust battery voltage for optimized pSCE operation.

Further, the optimal battery load $V^{II}_{pb,opt}$ corresponds to reducing the electrostatic energy loss during the second half-cycle (denoted $W_e^{II}$ above and marked as “Loss 1” in Fig.~\ref{Energy_Flow}) to zero.
Also, inclusion of the pre-biasing circuit's loss (“Loss 3” in Fig.~\ref{Energy_Flow}) and that of the extraction circuit (“Loss 2” in Fig.~\ref{Energy_Flow}) due to the resistivity of nonideal inductors involved in those circuits yields a more refined value of $V^{II}_{pb,opt}$. 
In particular, as shown in Supplementary Note~S.III\cite{pathak2021prebias}\cite{pathak2021prebias_arxiv}, factoring the inductor nonideality with quality factor 20, the revised $V^{II}_{pb,opt}$ value turns out to be 13.4$\%$ lower compared to the case of ideal inductor given by Eq.~(\ref{eq:C3}) (for our TENG implementation with parameters in Table~\ref{tab:Para}). Finally, the result remains robust to the change in the ambient vibration: For the same TENG parameters, a 20$\%$ change in the TENG amplitude $x_{max}$, changes the $V^{II}_{pb,opt}$ by less than 2$\%$.
\begin{figure}[htbp]
\vspace*{-.125in}
  \begin{center}
  \includegraphics[width=1.0\linewidth]{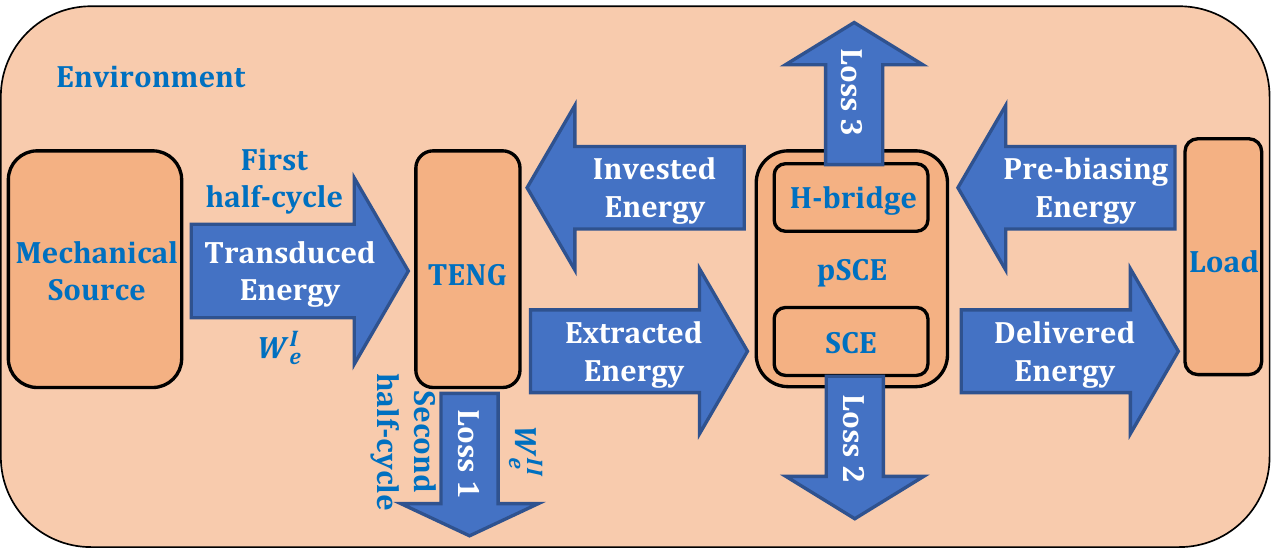}
  \vspace*{-.15in}
  \caption{Energy flow diagram for the pSCE operation. Loss 1 is electrostatic environmental loss while Loss 2 and 3 are parasitic resistive losses.}\label{Energy_Flow}
  \end{center}
\vspace*{-.075in}
\end{figure}

It follows that pre-biasing to $-2V_B$ in the second half cycle can in the best case be optimum (when $2V_B \geq V^{II}_{pb,opt}$), but otherwise it can be sub-optimal (when $V^{II}_{pb,opt}>2V_B$). We next show that even in the sub-optimal operation of the second half cycle (namely, pre-biasing TENG voltage to $-2V_B$ in all cases), the net energy gain of pSCE over SCE in the combined two cycles is positive for any $V_B$ value. This can be seen from the following sequence of equivalences:
\vspace{-0.05in}
\begin{align}
    &\left[E_{pSCE,net}|_{V^{II}_{pb}=2V_B}\right]-E_{SCE}\geq 0\nonumber \\
   \Leftrightarrow&(V_{oc,max}+2\beta V_B)^2+\frac{(V_{oc,max}+2V_B)^2}{\beta}\nonumber \\
   &-4(1+\beta) V_B^2-(1+\frac{1}{\beta})V_{oc,max}^2\geq 0\nonumber\\
    \Leftrightarrow&(\beta+\frac{1}{\beta})V_{oc,max}+(\beta^2+\frac{1}{\beta}-\beta-1)V_B\geq 0.\label{eq:C1}
\end{align}
\vspace{-0.05in}
By definition, $\beta \geq 1$, which implies that $(\beta^2+\frac{1}{\beta}-\beta-1)=(\beta-1)(\beta -\frac{1}{\beta})\geq0$. Also, the maximum open circuit voltage, $V_{oc,max}$ is a positive quantity. Thus, the condition of Eq.~(\ref{eq:C1}) is satisfied for all battery values for any given TENG. Also, in the final inequality of Eq.~(\ref{eq:C1}), the left hand side is an increasing function of $V_B$; the implication being that the energy gain of pSCE circuit over the SCE circuit shall continue to rise with increasing value of $V_B$. Thus, barring an eventual reduction in $x_{max}$ due to the increased pre-biasing or air electric field breakdown (as discussed earlier in {\em Remark~\ref{pSCE_EI}}), there is no other upper limit on $V_B$ as far as being able to boost the energy output through pre-biasing. Pre-biasing to a level higher than the presented design's upper limit of twice the battery voltage ($\pm 2V_B$) is also possible, but only at the added energy cost of DC/DC boosting. It is an easy exercise to check if the {\em net} energy gain by introducing such a boosted pre-biasing would be positive for a given TENG, and if so, a DC/DC boost can be easily integrated if the application can afford its added area.
\end{r1}
\begin{r1}
One may note that pSCE uses 4 inductors as opposed to 2 used by SCE. Then by choosing inductors of half the quality factor as that used in SCE, the pSCE output can be compared to that of SCE at identical inductor volume \cite{dicken2012power} (it is known that quality factor of inductors is proportional to its volume\cite{rand1963inductor}). It can be shown that even with half the quality factor inductors, pSCE provides higher energy output than SCE using inductors of quality factor as low as 4 using our implemented TENG's parameters listed in Table~\ref{tab:Para}. The derivation of the same is given in Supplementary Note~S.VI \cite{pathak2021prebias}\cite{pathak2021prebias_arxiv}. 
\end{r1}

\begin{figure*}
\vspace*{-.1in}
  \begin{center}
  \includegraphics[width=0.95\linewidth]{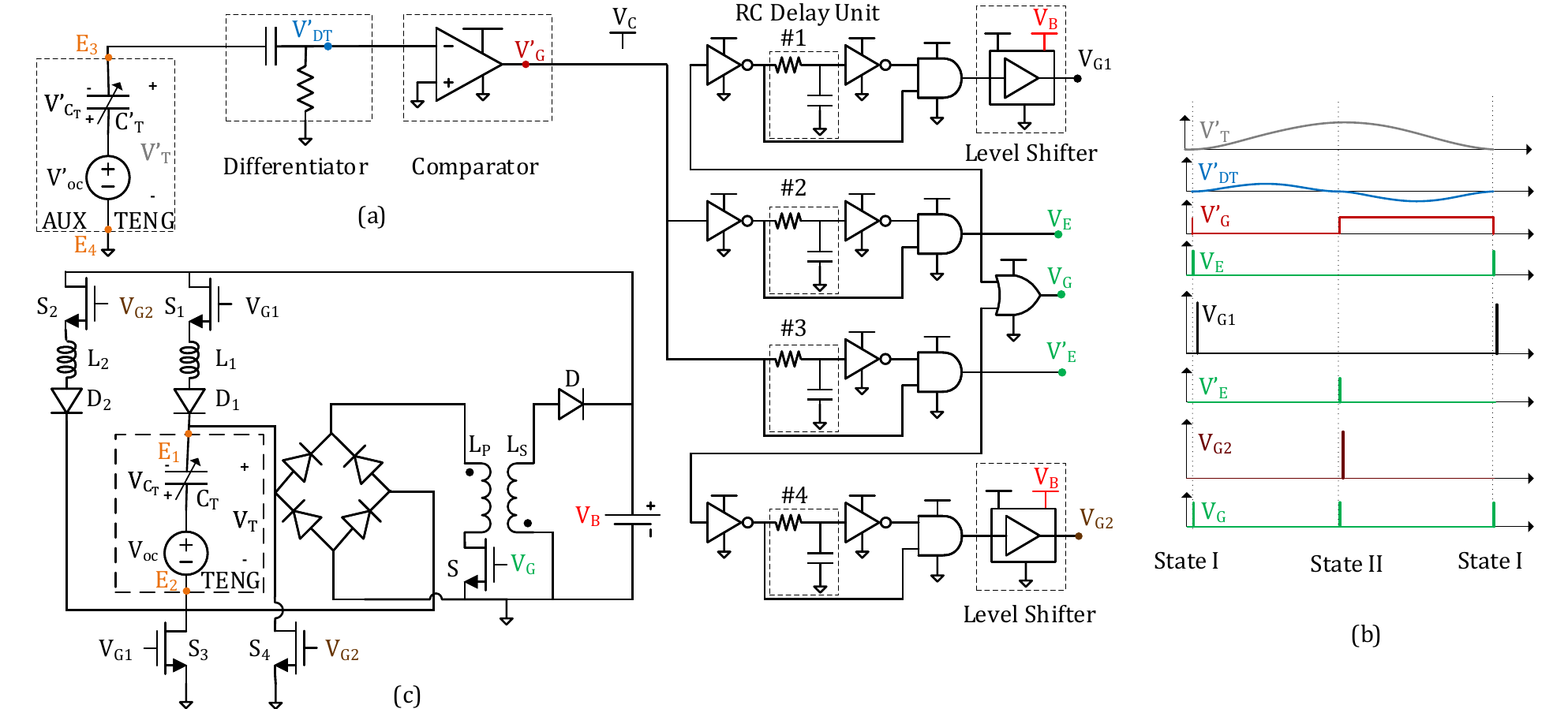}
  \vspace*{-.15in}
  \caption{(a) Control Circuit, (b) Its Waveforms and (c) Schematic for the pSCE circuit implementation.}\label{StripIII}
  \end{center}
\vspace*{-.3in}
\end{figure*}
\vspace*{-0.125in}
\section{Experimental Implementation}\label{Implementation}
\subsection{Experimental Setup}\label{A_Setup}
\vspace*{-.05in}
\begin{figure}
  \begin{center}
  \includegraphics[width=0.95\linewidth]{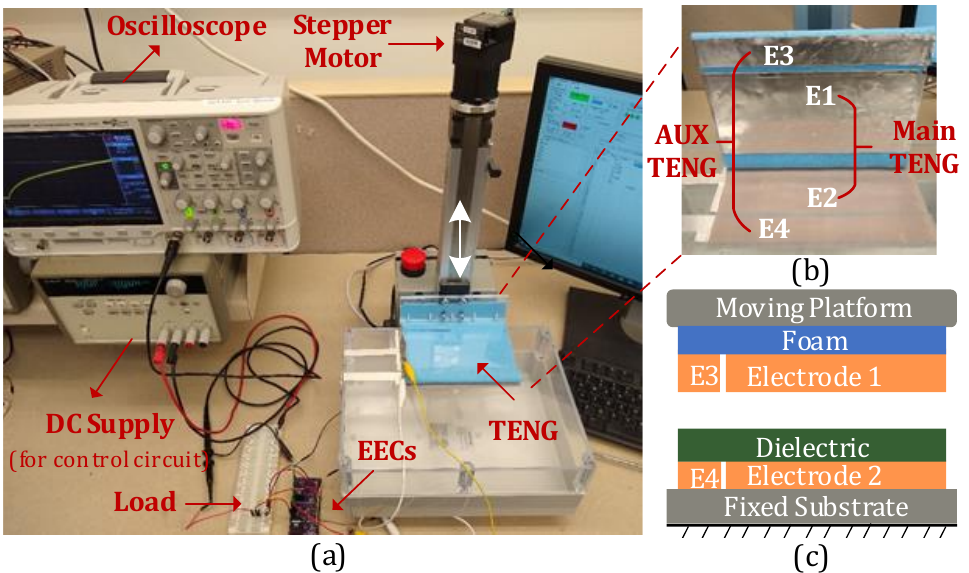}
  \vspace{-.15in}
  \caption{(a) Experimental Setup (b) Disassembled view of TENG (c) Schematic cross-section diagram of the implemented TENG.}\label{Setup}
  \end{center}
\vspace*{-.2in}
\end{figure}
\begin{figure}
  \begin{center}
  \includegraphics[width=0.95\linewidth]{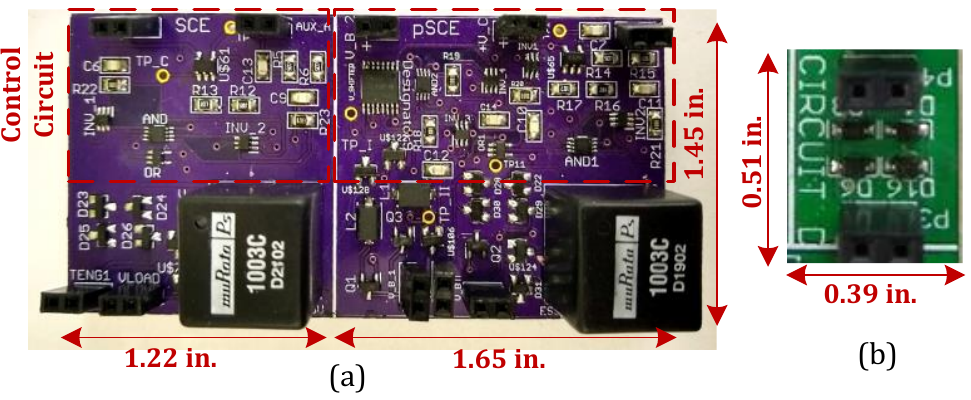}
  \vspace{-.15in}
  \caption{PCB implementations of the (a) SCE circuit, pSCE circuit, and (b) FWR circuit.}\label{PCB}
  \end{center}
\vspace*{-.3in}
\end{figure}

The overall experimental setup is shown in Fig.~\ref{Setup}(a). A computer programmable stepper motor is used to move the upper electrode plate relative to the fixed bottom electrode plate (a blow up of the same is shown in Fig.~\ref{Setup}(b)). Fig.~\ref{Setup}(a) also shows the implemented EEC and load. Output is measured using the oscilloscope and the data is collected through the NI’s Data Acquisition Card and Labview program. The DC supply powers the control circuit so that the energy consumed by the control circuits of SCE/pSCE can be measured non-interferingly. Fig.~\ref{Setup}(b) shows the main TENG (used for energy harvesting) in parallel with the auxiliary TENG (used for synchronously activating the control circuit without loading the main TENG). Fig.~\ref{Setup}(c) shows the schematic cross-section diagram of the TENG layers. The PCB implementations of the SCE and the pSCE circuits are shown in Fig.~\ref{PCB}(a), whereas that of FWR circuit is shown in Fig.~\ref{PCB}(b).

\vspace{-0.2in}
\subsection{TENG Setup}
TENG with a contact area of 112.5~cm$^2$ is experimentally implemented, as shown in the schematic of Fig.~\ref{StripI}(a). The fixed bottom plate is made of Teflon (dielectric) of thickness 127~$\mu$m on top of an Al sheet (electrode). The upper plate with the Al sheet (electrode) is driven in a reciprocating fashion using a stepper motor at 10 Hz frequency, with an amplitude ($x_{max}$) of  1.64~$mm$. An electrically isolated auxiliary TENG (Aux-TENG) with 1/5th the main TENG area is created to operate synchronously in parallel, and is tapped using Electrodes 3 and 4 to provide the input signal to the control circuit, issuing switching pulses to the SCE/pSCE circuits in synchrony with the main TENG. Independent Aux-TENG is required for distinguishing State I from State II from the measured signal and accordingly set the switching time $T_{SCE}^I$ or $T_{SCE}^{II}$ (Eq.~(\ref{eq:S5})) and additionally in case of pSCE circuit enable either pair of switches (S1,S3 or S2,S4) to pre-bias the TENG with the correct polarity. Use of Aux-TENG also avoids the loading of the main TENG by the control circuit and subsequent performance degradation.
\vspace*{-.025in}
\begin{table}[h]
\vspace*{-.1in}
    \centering
    \caption{Measured TENG Parameters}
    \label{tab:Para}
    \vspace*{-.05in}
    \begin{tabular}{c|c}
    \hline
    Maximum open circuit voltage: $V_{oc,max}$  & 279.92 V\\
    
    Minimum TENG capacitance: $C_{T,min}$ & 75.97 pF\\
    
    Maximum TENG capacitance: $C_{T,max}$ & 239.23 pF\\
    
    TENG capacitance ratio: $\beta$ & 3.15\\
    \hline
    \end{tabular}
    \vspace*{-0.2in}
\end{table}
\vspace*{-0.1in}
\subsection{TENG Characterization}
TENG is characterized by the three main parameters:  minimum and maximum TENG capacitance ($C_{T,min}$ and $C_{T,max}$), and maximum open circuit voltage ($V_{oc,max}$), that are summarized in Table~\ref{tab:Para} for our experimental setup. Dynamic variation of TENG capacitance is measured using the phase response based method described in \cite{basset2009batch,ghaffarinejad2018conditioning} and 
is plotted in Fig.~S4 of Supplementary Note~S.VII
from which the maximum and minimum values of the TENG capacitor are obtained. On the other hand, the maximum open circuit voltage is measured by charging a capacitor unto saturation with Full Wave Rectifier (FWR) circuit and using the below relation\cite{pathak2021synchronous},
\begin{equation*}
    V_{oc,max}=(\beta +1)(V_{sat}+2V_D).
\end{equation*}
The measured saturation voltage ($V_{sat}$) is $67.45V$. At saturation, negligible current flows through the FWR circuit and hence the diode drop ($V_D$) can be neglected, giving $V_{oc,max}\approx (\beta +1)V_{sat}=279.92 V$ for the experimentally determined $\beta=3.15$.  
\vspace*{-0.15in}
\subsection{Circuit Implementation}
\vspace*{-0.05in}
The pSCE circuit operates by switching at the extrema (State I and II), with the NMOS switches receiving the signal from the control circuit. Both the pSCE circuit and its control circuit are implemented as per the schematic diagram of Fig.~\ref{StripIII} using off-the-shelf components over a PCB as shown in the Fig.~\ref{PCB}(a).

The control circuit of Fig.~\ref{StripIII}(a) can be understood by following the voltage waveforms of Fig.~\ref{StripIII}(b). Its Boolean high output corresponds to $V_C=3V$ and low of $0V$. It receives the input signal from the Aux-TENG ($V'_T$), which gets differentiated by the RC circuit ($V'_{DT}$) to convert the signal peaks into zero crossings, which trigger state change of the comparator ($V'_G$) at States I and II. All four pulse generating arms of the circuit are designed to generate a pulse at the rising edge of the input: Arm 3 receives the comparator output and generates a pulse at State II ($V'_E$). Similarly, Arm 2 generates the pulse at State I ($V_E$) on receiving the inverted comparator output. The pulse width is set by the RC product of the respective delay unit. It is set to $T_{SCE}^I$ and $T_{SCE}^{II}$ (Eq.~(\ref{eq:S5})) for the Arms 2 and 3, respectively. Both these signals are combined  through OR gate to produce signal $V_G$ for the MOSFET S.  Arm 1 is used to generate a pulse for switching on MOSFET S1 and S3 ($V_{G1}$) at the falling edge of the signal $V_E$ to perform pre-biasing at State I+. The source of S1 reaches voltage upto $V_B$ by the end of the pre-biasing action (State I++), hence to turn it on, the signal ($V_{G1}$) is shifted up from $V_C$ to $V_B$ using the voltage level shifter. The pulse width of this signal should be greater than or equal to half the $L_1$-$C_{T,max}$ resonator cycle ($T_P^I$) to complete the pre-biasing action. At the end of the half-cycle, the current direction reverses and is automatically cut-off due to diode $D_1$ in the loop. Similarly, $V_{G2}$ is generated using arm 4 to switch on MOSFET S2 and S4 for pre-biasing at State II+. 

Additionally, SCE and the standard Full Wave Rectifier (FWR) circuits are implemented over PCB (shown in Fig.~\ref{PCB} (a) and (b), respectively.) for performance comparison with the presented pSCE circuit. Note the SCE circuit is implemented similar to the pSCE circuit discussed above, barring the H-bridge and the control circuit's arms 1 and 4 (refer Fig.~\ref{StripIII}). Finally, the FWR implementation is straightforward with the use of four diodes.
\vspace*{-0.15in}
\section{Results and Discussion}\label{Results}
\begin{figure*}
\vspace*{-.1in}
  \begin{center}
  \includegraphics[width=0.9\linewidth]{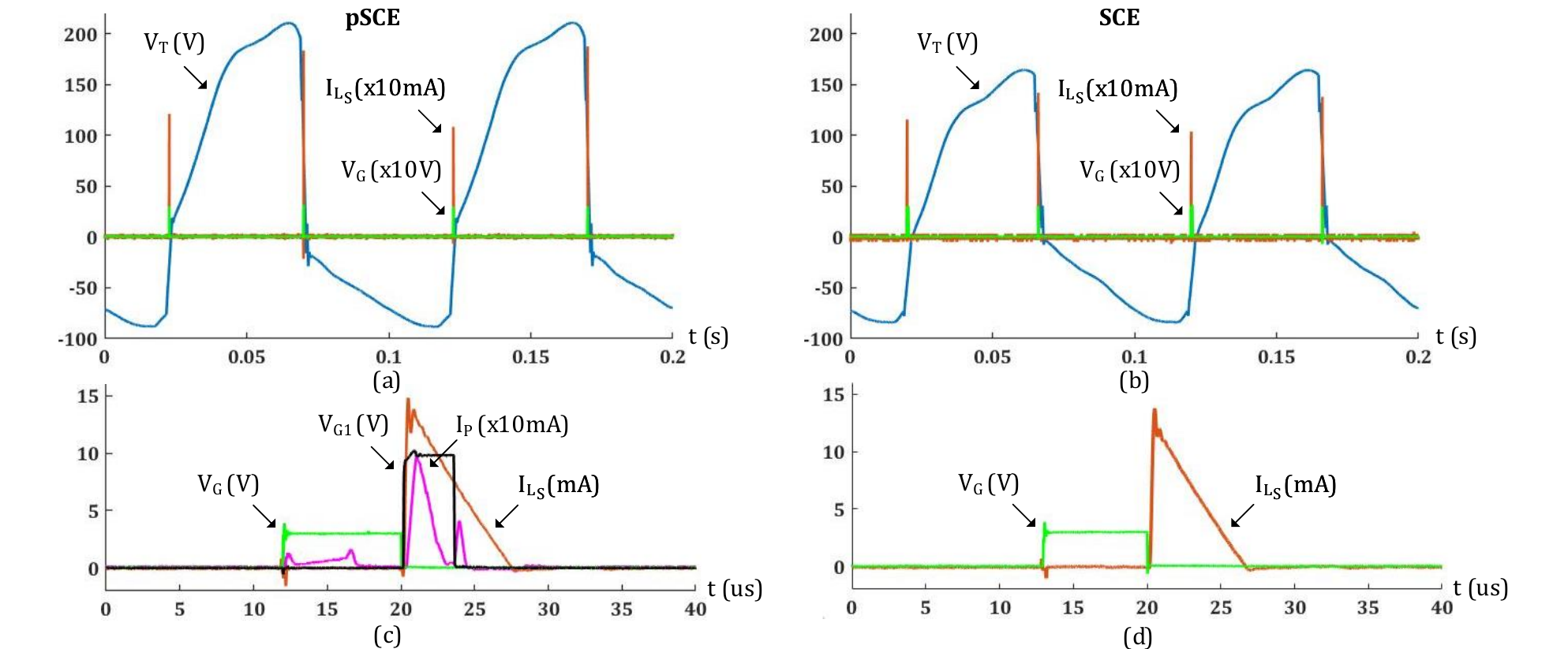}
  \vspace{-.05in}
  \caption{Measured TENG voltage ($V_T$), load current ($I_{L_S}$), and the control voltage ($V_{G}$) waveforms at 10V battery load for (a) pSCE circuit and (b) SCE circuit. (c) Zoomed-in view of $I_{L_S}$, $V_{G}$, pre-biasing control signal ($V_{G1}$) and pre-biasing current ($I_P$) for the pSCE circuit at State I. (d) Zoomed-in view of $I_{L_S}$ and $V_{G}$ for the SCE circuit at State I. }\label{Waveform}
  \end{center}
\vspace*{-.25in}
\end{figure*}

\vspace*{-0.05in}
Given $V_{oc,max}=279.92$ and $\beta=3.15$ (see Table~\ref{tab:Para}), $V^{II}_{pb,opt}=\frac{V_{oc,max}}{\beta -1}=130.19~V$. Since we employ battery load of at most $16~V$ for testing, in our experiments it holds that $2V_B\leq V^{II}_{pb,opt}$, and so following {\em Remark~\ref{pSCE_Opt}}, we pre-bias to $-2V_B$ at State II++. (The pre-biasing at State I++ is always at the maximum possible level of $2V_B$.) 
The measured TENG voltage ($V_T$), load current ($I_{L_S}$), and the control voltage ($V_{G}$) waveforms for the pSCE circuit for a $10~V$ battery load are shown in Fig.~\ref{Waveform}(a). As expected from our derivation, the TENG voltage, $V_T$ shows a pre-biasing voltage of value almost twice the battery load ($+/-\ 20~V$) post switching, i.e., at State I++ and State II++. Also, due to pre-biasing, the maximum TENG voltage in both the half-cycles, i.e., at State I and State II, is greater in value (in absolute terms) than its SCE counterpart (Fig.~\ref{Waveform}(b)). Fig.~\ref{Waveform}(c) shows the zoomed-in view of load current ($I_{L_S}$) and the control voltage ($V_{G}$) at State I. Fall of signal $V_{G}$ (MOSFET N turns off) triggers the rise of secondary inductor load current ($I_{L_S}$).  It additionally shows the measured pre-biasing control signal ($V_{G1}$) that is triggered by the fall of $V_{G}$ and the pre-biasing current ($I_{P}$) that flows through the $V_B-S_1-L_1-D_1-C_T-S_3$ path. The half-sinusoid $I_{P}$ current plot confirms pre-biasing for half the $L_1$-$C_{T,max}$ resonance cycle. Similarly, Fig.~\ref{Waveform}(d) shows the zoomed-in view of $I_{L_S}$ and $V_{G}$ at State I for the SCE circuit. Similar plots are observed at State II.
\vspace*{-.15in}
\subsection{Performance Comparison}
\begin{figure}[htbp]
  \begin{center}
  \includegraphics[width=0.95\linewidth]{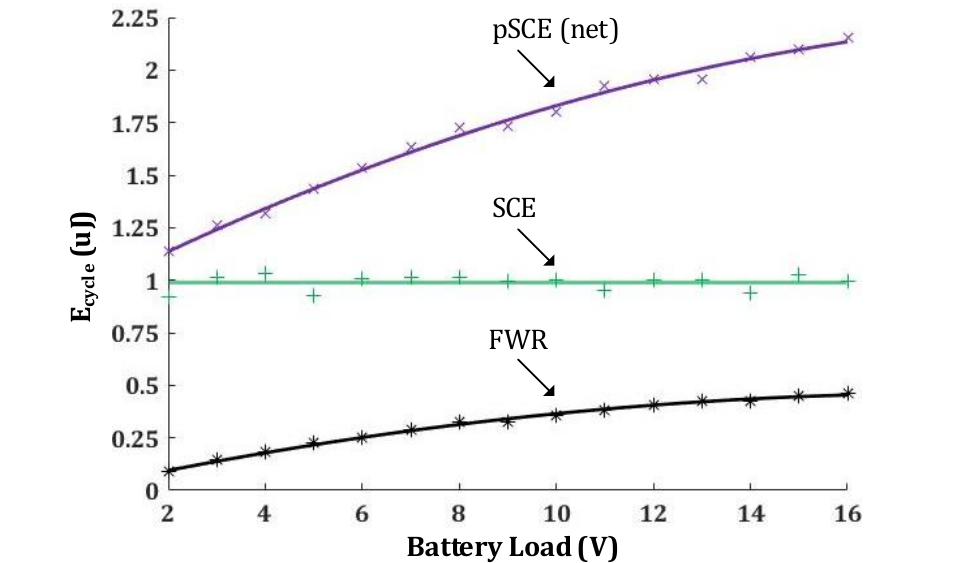}
  \vspace{-.05in}
  \caption{Comparison of measured energy per-cycle for FWR, SCE, and pSCE circuit against the battery load.}\label{E}
  \end{center}
\vspace*{-.4in}
\end{figure}

\vspace*{-.05in}
The fundamental comparison of the first time presented pSCE circuit of this work is with its peer SCE circuit, together with the base case FWR circuit. For our experimental TENG with the parameters listed in Table~\ref{tab:Para}, the measured energy output per-cycle ($E_{cycle}$) using FWR, SCE, and pSCE circuit as Energy Extraction Circuits (EECs) against battery loads are compared in the plot of Fig.~\ref{E}. It is calculated as the product of load battery ($V_B$) and integration of the measured current (charge) flowing through the load battery over one cycle. For the pSCE circuit, the net energy per-cycle ($E_{pSCE,net}$) is plotted, which is obtained by deducting the pre-biasing energy ($E_{pre-bias}$ (calculated as the product of $V_B$ and the integration of pre-biasing current, $I_P$ over one cycle) from the gross output energy ($E_{pSCE}$).
\begin{table}[htbp]
\vspace*{-0.05in}
\centering
\caption{Measured energy and gain of the pSCE circuit over SCE and FWR circuit at different battery load.}\label{tab:E}
\vspace*{-0.075in}
\begin{adjustbox}{width=1.0\linewidth}
\begin{tabular}{c|c|c|c|c|c}
\hline
\thead{$V_B$\\(V)}
 & \thead{$E_{pSCE}$\\(uJ)}  & \thead{$E_{pre-bias}$\\(uJ)}   & \thead{$E_{pSCE,net}$\\(uJ)}  & \thead{Gain\\over SCE} & \thead{Gain\\over FWR} \\ \hline
5                                                            & 1.455                                                              & 0.018                                                                  & 1.437                                                                  & 1.453                                                             & 6.653                                                                 \\ 
10                                                            & 1.853                                                              & 0.052                                                                  & 1.801                                                                  & 1.821                                                             & 4.934                                                                 \\ 
15                                                           & 2.185                                                              & 0.088                                                                  & 2.097                                                                  & 2.120                                                              & 4.705                                                                 \\ \hline
\end{tabular}
\end{adjustbox}
\vspace*{-0.1in}
\end{table}

As evident from Fig.~\ref{E}, $E_{cycle}$ for FWR shows the expected parabolic response with variation in battery load\cite{niu2015optimization,pathak2021synchronous}. SCE circuit has a constant energy output irrespective of the battery load value ($V_B$) as expected from the circuit analysis in Sec.~\ref{Sec_SCE} and confirming {\em Remark~\ref{Passive_EEC}}. For the pSCE circuit, an increasing net energy output trend is observed with the increasing value of $V_B$ validating the theoretical development in Sec.~\ref{Sec_pSCE}. As $V_B$ increases, the TENG voltage at State II ($V_T^{II}$) increases as $V_{oc,max}+2\beta V_B$. The MOSFET switches used in our implementation start to saturate at around $V_B$ of 15V which is reflected in the $E_{pSCE,net}$ plot of Fig.~\ref{E}. 
\begin{table}[htbp]
\vspace*{-.1in}
\centering
\caption{Experimental Power Density of the FWR, SCE, and the pSCE circuits at 10V battery load.}\label{tab:Power_Density}
\vspace*{-0.075in}
\begin{threeparttable}[h]
\begin{tabular}{c|c|c|c|c}
\hline
\multirow{2}{*}{\begin{tabular}[c]{@{}c@{}} Circuit \\\end{tabular}} & \multirow{2}{*}{\begin{tabular}[c]{@{}c@{}} PCB Area \\(cm$^2$)\end{tabular}} & \multirow{2}{*}{\begin{tabular}[c]{@{}c@{}}Power Density\\(uW/cm$^2$)\end{tabular}} & \multirow{2}{*}{\begin{tabular}[c]{@{}c@{}}PCB+TENG\\Area (cm$^2$)\end{tabular}} & \multirow{2}{*}{\begin{tabular}[c]{@{}c@{}}Power Density\\ (uW/cm$^2$)\end{tabular}} \\
&   &  &   & \\
\hline
FWR                      & 1.290                             & 2.829                          & 113.790                                                                                            & 0.032                                                                            \\
SCE                      & 11.413                                                                                 & 0.867                                                                               & 146.413\tnote{a}   
& 0.068   
\\
pSCE                     & 15.435                                                                                 & 1.167                                                                               & 150.435\tnote{a}   
& 0.120 
\\
\hline
\end{tabular}
        [a] includes Aux-TENG area
    \end{threeparttable}
\vspace*{-0.1in}
\end{table}
Table~\ref{tab:E} lists the gain of pSCE circuit output over the state-of-art SCE circuit and the standard FWR circuit for three different battery load. The pSCE circuit's gain increases with higher pre-biasing through the battery as the SCE output remains constant. Initially, the gain over FWR decreases with rising $V_B$ due to the parabolic nature of FWR's $E_{cycle}$ but is expected to rise later at higher $V_B$. Further, the power density figures of FWR, SCE, and pSCE circuits from our experimental implementation at 10 Hz operational frequency and 10V battery load are listed in Table~\ref{tab:Power_Density}. It should be noted that we implement the TENG only for circuit validation using readily available material such as Aluminum and Teflon tape. The power density values can significantly improve in practical implementation using the advances in TENG materials and designs.

\begin{table}[htbp]
\vspace*{-.1in}
    \centering
    \caption{Performance comparison with reported EECs for TENG}
    \label{tab:benchmark}
    \begin{adjustbox}{width=1.0\linewidth}
    \begin{threeparttable}[h]
    \begin{tabular}{c|c|c|c|c}
       \hline
       Ref  & \thead{Circuit\\Architecture}   &  \thead{Maximum theoretical\\efficiency ($\eta^*$)\\(and comment)} & \thead{Measurement\\Load} & \thead{Experimental\\efficiency ($\eta$)\tnote{a}}  \\
       \hline
      \cite{pathak2021synchronous}     & FWR   & \thead{$\leq$~25$\%$\\(max at $\beta=1$)}     &\thead{26.75~V\\(Optimal)}    & 20.43$\%$  \\
       \thead{This\\Work}   & FWR   & $\leq$~25$\%$    & 15~V  & 11.37$\%$ \\
       \cite{ghaffarinejad2018conditioning}     & \thead{Half wave\\rectifier with\\parallel diode}  & 25$\%$   & $\sim$ 58~V    &15.49$\%$\\
       \cite{zi2017inductor}    & \thead{Synchronous\\switched\\serial/parallel\\capacitor}     &\thead{25$\%$\\(depends on load \&\\number of intermediate\\capacitors)}    &$\sim$90~V   &\thead{23$\%$\\(mechanical\\switching)}\\
       \cite{zi2016effective}   & \thead{Synchronous\\parallel switch}  &\thead{$\leq$~50$\%$\\(max at $\beta=1$)}     &\thead{$\sim$70~V\\(Optimal)}  &\thead{$\sim$42.86$\%$\\(mechanical\\switching)} \\
       \cite{ghaffarinejad2018conditioning}     & \thead{Bennet voltage\\doubler}   &\thead{Unbounded; limited\\by air breakdown\\(needs $\beta>2$)} &$\sim$50~V    &58.43$\%$ \\
       \cite{pathak2021synchronous}     &P-SSHI    &\thead{Limited by quality\\factor; high optimal\\load, so, impractical}   &15~V   &19.53$\%$ \\
       \cite{pathak2021synchronous}     &S-SSHI    &\thead{Limited by quality\\factor; multiple\\transient cycles}  &\thead{26.15~V\\(Optimal)}  &172.8$\%$ \\
       \cite{cheng2017high} &SCE    &\thead{100$\%$\\(capacitor load)}  &-  &29.6$\%$\\
       \cite{wu2020high} &SCE    &\thead{100$\%$\\(capacitor load)}  &-  &37.8$\%$\\
       \thead{This\\Work} &SCE    &\thead{100$\%$\\(battery load)}  &-  &25.22$\%$\\
       \thead{This\\Work} &pSCE    &\thead{Unbounded; limited\\by air breakdown\\(161.72$\%$ at 15~V)}  &15~V  &53.48$\%$\\
    
    \hline
    \end{tabular}
        [a] ignores control circuit energy consumption
\end{threeparttable}
\end{adjustbox}
\vspace*{-0.25in}
\end{table}

\begin{r1}
For comparing the conversion efficiency of the presented circuits with those reported in literature, the Cycle for Maximized Energy Output (CMEO) \cite{zi2015standards} serves as the reference and has also been used to calculate efficiency in previous works such as \cite{zi2017inductor} and \cite{zi2016effective}. As mentioned in Introduction, CMEO uses simple switching at the two extremes to let the charge flow from one plate to the other via an ideal infinite load, thereby supplying energy to it; the TENG's corresponding voltage-vs-charge diagram yields per-cycle energy as: $E_{CMEO}=\frac{1}{2}\left(1+\frac{1}{\beta}\right)C_{T,min}V_{oc,max}^2$ \cite{zi2015standards}. The energy conversion efficiency with reference to CMEO is then calculated as: $\eta=\frac{E_{cycle}}{E_{CMEO}}\times100\%$. Table~\ref{tab:benchmark} first lists the maximum theoretical {\em ideal-circuit} efficiency ($\eta^*$) that can possibly be achieved at their corresponding optimal loads. Also, as noted in {\em Remark~\ref{Passive_EEC}}, SCE theoretically attains $E_{CMEO}$ ($\eta^*=100\%$), while, in contrast, the proposed pSCE always outperforms SCE and its energy output continues to grow with the load voltage and hence its $\eta^*$ is unbounded (Refer {\em Remark~\ref{pSCE_Opt}}). Table~\ref{tab:benchmark} next lists the reported experimental load voltage and the measured efficiency $\eta$ at that voltage. The energy outputs of FWR, S-SSHI, P-SHHI and other circuits listed in the table depend on load voltage, and an optimized load is needed to maximize the energy output, which requires a DC/DC converter and consumes additional energy, and will further reduce the net energy output. A key advantage of the proposed pSCE architecture is that it offers energy output greater than CMEO at any load battery voltage as derived in Eq.~(\ref{eq:C1}).

\vspace*{-0.2in}
\subsection{Overheads and non-idealities}
\vspace{-0.025in}
Both SCE and pSCE require overhead power for the associated control circuit: The per-cycle control circuit consumption for SCE and pSCE circuit were measured as $0.293\ uJ$ and $0.423\ uJ$ (at $V_B=10V$), respectively. It should be noted that for our validation experiments, off-the-shelf components are used to implement the control circuit as a proof-of-study. Hence, there is a large room for further optimization to reduce this consumption, for example, by using a custom designed IC. 

Note both pre-biasing and control circuits require power to operate, and to achieve a cold start in a practical implementation; load battery would need to be charged first by operating in the FWR mode to a level that it can supply the needed power for the SCE control circuit, and start the operation in the SCE mode. The pSCE action would start later when the battery voltage rises above the H-bridge path's threshold voltage of one diode and two switches, typically $\sim$2V. 

We can also observe the role of non-idealities from Table~\ref{tab:benchmark}, where for both the SCE and pSCE circuits, their efficiencies of 25.22$\%$ and 53.48$\%$, resp. are lower than their theoretical values of 100$\%$ and 161.72$\%$, resp.: For example, both the SCE and pSCE circuits operate with switches off most of the time (barring the short durations at States I and II), and the leakage of charge through the non-ideal MOSFET switches during the off periods results in lower voltage magnitudes at States I and II and hence, lower extracted energy. 
The inductors ($L_1$ and $L_2$) used for pre-biasing have finite quality factors, and hence the pre-biasing voltage falls short of $2V_B$, in turn leading to a lower energy output than the theoretically expected value. Another practical issue is the delay in the control circuit, which leads to switching past the peak at States I and II as observed from the voltage waveforms of Fig.~\ref{Waveform}(a) and \ref{Waveform}(b). Other non-idealities include diode voltage drops, on-resistance of MOSFET switches, and the coupled inductors' parasitics.
\end{r1}

\vspace*{-0.15in}
\section{Conclusion}
This work proposed {\em active} pre-biasing of Triboelectric Nanogenerator (TENG) using its already present load battery for boosting the output energy, even beyond the Synchronous Charge Extraction (SCE) architecture, one that is proven to operate at CMEO (cycle for maximized energy output), but in a {\em passive} setting. It was shown that the increase in output energy due to pre-biasing in the first half-cycle of TENG operation (separation of the TENG plates) is attributed to the increase in the transduced energy from the mechanical source due to the addition of extra charges on the plates, leading to more mechanical work against the induced electrostatic attraction. In contrast, the gain in the second half-cycle (retraction of plates) is due to a reduction in the TENG electrical (potential) energy that is dissipated back into the environment (retraction motion being in the same direction as electrostatic attraction). We showed that this loss can be reduced to zero by using pre-biasing to set the plate charges to zero, and analytically showed that this is indeed the optimum from the energy perspective. It was further shown that increasing the pre-biasing increases the overall output, and so in principle the mechanical energy can be fully harvested. The upper limit to pre-biasing is essentially determined by the level of available mechanical excitation (acceleration), namely, till the electrostatic attraction due to pre-biasing become comparable to the external excitation force. For the second half-cycle, the optimum pre-biasing voltage was derived to cut the energy loss to zero, and also the upper limit for pre-biasing in the second half cycle was derived beyond which the pSCE will have higher losses than SCE in the second half cycle.

For the implementation of the proposed technique, Pre-biased Synchronous Charge Extraction circuit (pSCE) circuit was presented to enable the pre-biasing of TENG in each half-cycle using the load battery. Pre-biasing voltage was passively boosted to 2x the battery voltage by forming an LC circuit with the TENG capacitor and an external inductor. The energy output of the pSCE circuit and conditions on pre-biasing voltages for net-benefit over the SCE circuit were derived. Experimental implementation of the pSCE circuit validates the expected gain in the energy output. Using the pSCE circuit for $5V$ battery load, gains of 1.453 over the SCE circuit and 6.653 over the standard Full Wave Rectifier (FWR) circuit were achieved. A future research direction could explore the use of low-power active DC-DC converters to achieve optimum pre-biasing. We believe that the presented effort to increase the energy output by designing a novel pSCE circuit will bring us closer to the real-world feasibility of powering wireless sensor nodes by TENG.




%




\ifCLASSOPTIONcaptionsoff
  \newpage
\fi




\vspace*{-0.175in}
\bibliographystyle{IEEEtran}
\bibliography{IEEEabrv,Bibliography}
\vfill

\ifarXiv
    \foreach \x in {1,2,3,4,5}
    {

\section*{Supplementary material}

# Supplementary Document: Synchronous Pre-biasing of Triboelectric Nanogenerator for Enhanced Energy Extraction

Madhav Pathak and Ratnesh Kumar, *Fellow, IEEE*

SUPPLEMENTARY NOTES:

S.I. DERIVATION OF PER-CYCLE ENERGY OUTPUT OF SCE WITH CONSIDERATION FOR INDUCTOR RESISTANCE LOSS

The energy output ($E_{SCE}$) derived in Sec. III-B through the $V_T$ vs. $Q_{C_T}$ plot represents the energy extracted from the source (TENG). Here, instead we derive the energy delivered to the load to include the inductor resistance loss in the extraction circuit.

The SCE circuit with switch S on at State I is simplified to that shown in Fig. S1(a). Here, $R_S$ is the series parasitic resistance of the inductor. The differential equation for the current in the loop can be written as:

$$L_P \frac{d^2 I_{L_P}^I(t)}{dt^2} + R_S \frac{d I_{L_P}^I(t)}{dt} + \frac{1}{C_{T,max}} I_{L_P}^I(t) = 0.$$

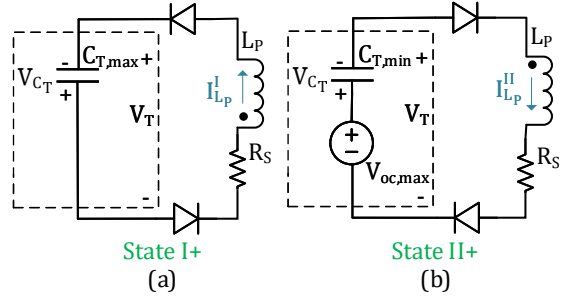

Fig. S1. Simplified circuit during energy extraction with Switch S on at (a) State I and (b) State II.

Note that the circuit loop with $L_P$ can be independently analyzed as both primary and secondary inductors are decoupled. The above equation can be solved with initial condition as $I_{L_P}^I(0) = 0$ and $\frac{d I_{L_P}^I(t)}{dt}(0) = \frac{V_{C_T}^I}{L_P} = \frac{V_{oc,max}}{\beta L_P}$ (Refer Eq. (6)) to obtain:

$$I_{L_P}^I(t) = \left(\frac{e^{-\frac{R_S t}{2 L_P}}}{L_P \omega_{d,SCE}^I}\right)\left(\frac{V_{oc,max}}{\beta}\right) sin(\omega_{d,SCE}^I t); \quad \omega_{d,SCE}^I := \sqrt{\frac{1}{L_P C_{T,max}} - \frac{R_S^2}{4 L_P^2}}.$$

The switch is opened at one-fourth the $L_P - C_{T,max}$ cycle with the peak current as,

$$I_{L_P}^I\left(\frac{\pi}{2\omega_{d,SCE}^I}\right) = I_{L_P,max}^I = \left(\frac{\sqrt{\alpha^I}}{L_P \omega_{d,SCE}^I}\right)\left(\frac{V_{oc,max}}{\beta}\right); \quad \alpha^I := e^{\frac{-\pi R_S}{2\omega_{d,SCE}^I L_P}} = e^{\frac{-\pi}{2 Q_f^I}}; \quad Q_f^I := \frac{\omega_{d,SCE}^I L_P}{R_S}. \quad (S1)$$

Here, $Q_f^I$ is the series resonator quality factor at State I, and $0 < \alpha^I < 1$ is its "normalized" form that effectively captures the resistive circuit loss in the extraction step. Opening switch $S$ leads to a sharp cut-off of $I_{L_P}^I$ in the primary loop. The negative voltage developed on the coupled inductors due to this current fall is conducive for current flow in the secondary side, and $I_{L_S}^I$ rises to a maximum of $I_{L_P,max}^I$ (taking coupling coefficient as 1 with $L_P = L_S$), which then is discharged through the diode $D$ into the load battery. Using KVL in the secondary loop gives,

$$L_S \frac{d I_{L_S}^{II}(t)}{dt} + V_B = 0 \Leftrightarrow I_{L_S}^{II}(t) = -\frac{V_B}{L_S} t.$$

Here, by nature of the circuit, DC parasitic resistance of the secondary inductor $L_S$ comes into play which is typically negligible and hence has been ignored in above equation. $I_{L_S}^I$ decays linearly from peak value of $I_{L_P,max}^I$ to zero during the charging duration, say $T_L^I$ (Refer Fig. 3(c)), which can be computed by integrating the above equation:

$$\int_{I_{L_P,max}^I}^0 d I_{L_S}^I(t) = \int_0^{T_L^I}\left(-\frac{V_B}{L_S}\right) dt \Rightarrow T_L^I = \frac{I_{L_P,max}^I L_S}{V_B}. \quad (S2)$$

The energy delivered to the battery load is total charge flowing over the time $T_L^I$ times the battery voltage which simplifies to,

$$E_{SCE}^{II} = V_B \int_0^{T_L^I} I_{L_S}^I(t)dt = V_B \int_0^{T_L^I} \left(-\frac{V_B}{L_S}t\right)dt = \frac{1}{2}L_P \left(I_{L_P,max}^I\right)^2. \tag{S3}$$

Substituting the value of $I_{L_P,max}^I$ from Eq. (S1) gives,

$$E_{SCE}^{II} = \frac{1}{2}\left(\frac{\alpha^I}{\beta}\right)C_{T,min}V_{oc,max}^2. \tag{S4}$$

Similar analysis at State II yields,

$$E_{SCE}^I = \frac{1}{2}\alpha^{II} C_{T,min} V_{oc,max}^2; \quad \alpha^{II} := e^{\frac{-\pi R_S}{2\omega_{d,SCE}^{II} L_P}} = e^{\frac{-\pi}{2Q_f^{II}}}; \quad Q_f^{II} := \frac{\omega_{d,SCE}^{II} L_P}{R_S}; \quad \omega_{d,SCE}^{II} := \sqrt{\frac{1}{L_P C_{T,min}} - \frac{R_S^2}{4L_P^2}}, \tag{S5}$$

where, $Q_f^{II}$, the series resonator quality factor at State II and its normalized form $0 < \alpha^{II} < 1$ are different than their State I counterparts (Eq. S1) due to change in TENG capacitance from $C_{T,max}$ to $C_{T,min}$. Then, the per-cycle energy delivered to the load shall be,

$$E_{SCE} = E_{SCE}^I + E_{SCE}^{II} = \frac{1}{2}\left(\frac{\alpha^I}{\beta} + \alpha^{II}\right)C_{T,min}V_{oc,max}^2 \tag{S6}$$

### S.II. Derivation of net per-cycle energy output of pSCE with consideration for inductor resistance loss

The SCE circuit acts as the extraction circuit in the pSCE operation. Hence, from Eq. (21) and following the derivation of the SCE circuit's output in Supplementary Note S.I above, the energy output of the pSCE circuit at the end of first half-cycle ($E_{pSCE}^I$) and that at the end of second half-cycle ($E_{pSCE}^{II}$) are given by,

$$E_{pSCE}^I = \frac{1}{2}(\alpha^{II})C_{T,min}\left(V_{oc,max} + \beta V_{pb}^I\right)^2; \quad E_{pSCE}^{II} = \frac{1}{2}\left(\frac{\alpha^I}{\beta}\right)C_{T,min}\left(V_{oc,max} + V_{pb}^{II}\right)^2, \tag{S7}$$

where, $V_{pb}^I$ and $V_{pb}^{II}$ are the pre-biasing voltages at States I and II, respectively. The normalized quality factors: $0 < \alpha^I, \alpha^{II} < 1$ at States I and II have been defined in Eq. (S1) and Eq. (S5), respectively.

Next, we analyze the H-bridge circuit considering resistive loss of the non-ideal inductors: $L_1$ and $L_2$ ("Loss 3" in Fig. 8) to derive the pre-biasing voltages ($V_{pb}^I$ and $V_{pb}^{II}$), and the energy used (invested) for the same ($E_{pre-bias}^I$ and $E_{pre-bias}^{II}$) at States I and II. For analytical convenience, we consider the case where TENG is pre-biased for half the $LC_T$ resonator cycle i.e., upto the maximum achievable voltage at both States I and II. First, we consider the pre-biasing circuit at State I++ with the switches $S_1$ and $S_3$ (Refer Fig. 6) closed. The differential equation for the pre-biasing loop is given by,

$$\frac{d^2V_{C_T}(t)}{dt^2} + \frac{R_S}{L_1}\frac{dV_{C_T}(t)}{dt} + \frac{V_{C_T}(t)}{L_1 C_{T,max}} - \frac{V_B}{L_1 C_{T,max}} = 0.$$

Solving the above equation for $V_{C_T}(t)$ with the initial condition as $V_{C_T}(0) = V_{C_T}^{I+} = 0$ and taking $L_1 = L_P$,

$$V_{C_T}(t) = e^{-\frac{R_S t}{2L_P}}(V_B)\left[\cos(\omega_{d,SCE}^I t) + \frac{R_S}{2L_P \omega_{d,SCE}^I}\sin(\omega_{d,SCE}^I t)\right] - V_B,$$

where, $\omega_{d,SCE}^I$ is same as defined earlier in Eq. (S1). With the switches S1 and S3 closed for half the $L_1 - C_{T,max}$ resonator cycle,

$$V_{C_T}\left(\frac{\pi}{\omega_{d,SCE}^I}\right) = -(1+\alpha^I)V_B.$$

Thus, a pre-biasing charge equal to $C_{T,max}(1+\alpha^I)V_B$ is added by the load battery corresponding to energy of

$$E_{pre-bias}^I = \beta C_{T,min}(1+\alpha^I)V_B \times V_B = (1+\alpha^I)C_{T,min}\beta V_B^2, \tag{S8}$$

and the pre-biasing voltage equal to $V_{pb}^I = V_T^{I++} = -V_{C_T}^{I++} = (1+\alpha^I)V_B$. Now, using Eq. (S7) and Eq. (S8), the net energy delivered at the end of first half-cycle can be found as,

$$E_{pSCE,net}^I = E_{pSCE}^I - E_{pre-bias}^I = \frac{1}{2}C_{T,min}\left[(\alpha^{II})\left(V_{oc,max} + (1+\alpha^I)\beta V_B\right)^2 - 2(1+\alpha^I)\beta V_B^2\right]. \tag{S9}$$

Performing similar analysis at the end of second half cycle gives $V_{pb}^{II} = (1+\alpha^{II})V_B$ and the net per-cycle energy output as,

$$E_{pSCE,net}^{II} = E_{pSCE}^{II} - E_{pre-bias}^{II} = \frac{1}{2}C_{T,min}\left[\frac{(\alpha^I)\left(V_{oc,max}+(1+\alpha^{II})V_B\right)^2}{\beta} - 2(1+\alpha^{II})V_B^2\right]. \quad \text{(S10)}$$

Then the net per-cycle energy output of the pSCE circuit is given by,

$$\begin{aligned}E_{pSCE,net} =& E_{pSCE,net}^I + E_{pSCE,net}^{II} \\ =& \frac{1}{2}C_{T,min}\left[(\alpha^{II})\left(V_{oc,max}+(1+\alpha^I)\beta V_B\right)^2 + \frac{(\alpha^I)\left(V_{oc,max}+(1+\alpha^{II})V_B\right)^2}{\beta} - 2\left(1+\alpha^{II}+(1+\alpha^I)\beta\right)V_B^2\right].\end{aligned} \quad \text{(S11)}$$

### S.III. DERIVATION OF OPTIMUM PRE-BIASING VOLTAGE WITH CONSIDERATION FOR INDUCTOR RESISTANCE LOSS

As discussed in Sec. V-C, the optimum pre-biasing voltage for the second half-cycle, $V_{pb,opt}^{II}$ is obtained at maximum difference between the pSCE output ($E_{pSCE,net}^{II}$) and SCE output ($E_{SCE}^{II}$). Using Eq. (S4) and substituting $V_B = \frac{V_{pb}^{II}}{(1+\alpha^{II})}$ in Eq. (S10), we obtain,

$$\begin{aligned}E_{pSCE,net}^{II} - E_{SCE}^{II} =& \frac{1}{2}C_{T,min}\left[\frac{\alpha^I\left(V_{oc,max}+V_{pb}^{II}\right)^2}{\beta} - 2\frac{\left(V_{pb}^{II}\right)^2}{(1+\alpha^{II})} - \frac{\alpha^I V_{oc,max}^2}{\beta}\right] \\ =& \frac{1}{2}C_{T,min}\left[\left(\frac{\alpha^I}{\beta}\right)\left(V_{pb}^{II}\right)\left(2V_{oc,max}+V_{pb}^{II}\right) - 2\frac{\left(V_{pb}^{II}\right)^2}{(1+\alpha^{II})}\right].\end{aligned} \quad \text{(S12)}$$

Differentiating above equation with respect to $V_{pb}^{II}$ and setting it equal to zero yields,

$$V_{pb,opt}^{II} = \frac{\alpha^I(1+\alpha^{II})V_{oc,max}}{(2\beta-\alpha^I-\alpha^I\alpha^{II})} \quad \text{(S13)}$$

With the minimum quality factor of the used inductor $L_2$ (CC453232-101KL) being $Q_f^I = 20$ [1] ($Q_f^{II} > Q_f^I$ since $\omega_{d,SCE}^{II} > \omega_{d,SCE}^I$), substituting the values TENG parameters, $V_{oc,max}$ and $\beta$ from Table I in above equation yields $V_{pb,opt}^{II} = 112.75\ V$, which is 13.4% lower than that obtained using $V_{pb,opt}^{II} = \frac{V_{oc,max}}{\beta-1}$ (Eq. (25)). The first-order formulation of Eq. (25) is independent of the operation parameter such as amplitude ($x_{max}$) and the revised formulation of Eq. (S13) is also found robust to the change in $x_{max}$. For example, a 20% change $x_{max}$ changes $V_{pb,opt}^{II}$ by less than 2%.

### S.IV. CALCULATION FOR ELECTROSTATIC ATTRACTION FOR SCE AND pSCE

Here, we perform first order calculations to determine the deceleration acting on the moving upper plate of TENG due to electrostatic force of attraction in the first half-cycle. For the SCE circuit, using Eq. (3), Eq. (10) and, Eq (11),

$$|\vec{F}_e^I| = \frac{(Q_1^{I+})^2}{2A\epsilon_0} \Rightarrow a_e^I = \frac{(Q_1^{I+})^2}{2mA\epsilon_0} = \frac{\sigma^2 A}{2m\epsilon_0}; \quad \sigma = \frac{V_{oc,max}\epsilon_0}{x_{max}}.$$

Here, $m$ is the mass of the upper plate (113.7$g$). Using the operation and TENG parameters (Refer Sec. VI), electrostatic deceleration ($a_e^I$) is calculated as 0.013 $m/s^2$. Similarly for the pSCE circuit (Refer Fig. 5(c)),

$$|\vec{F}_e^I| = \frac{(Q_1^{I++})^2}{2A\epsilon_0} \Rightarrow a_e^I = \frac{(Q_1^{I++})^2}{2mA\epsilon_0} = \frac{(\sigma A + 2C_{T,max}V_B)^2}{2mA\epsilon_0}.$$

The above electrostatic attraction is calculated as 0.026 $m/s^2$ at battery load ($V_B$) of 15$V$. This is at least 2 orders of magnitude smaller compared to acceleration of natural sources (such as 2-3 $m/s^2$ for human walking, while 12 $m/s^2$ for car compartment [2]).

## S.V. Lossless pre-biasing at State II++ for arbitrary voltage

As discussed in Sec. V-A, at the start of the second half cycle (State II++), the TENG voltage can be set to any value $-V_{pb}^{II}$ such that $-2V_B \leq -V_{pb}^{II} \leq 0$ by adjusting the duration of pre-biasing. If the switches $S_2$ and $S_4$ are enabled for half the $L_2 - C_{T,min}$ resonator cycle, the maximum value of $V_{pb}^{II} = 2V_B$ is obtained. By reducing the "on" duration of the these switches, a lower $V_{pb}^{II}$ voltage can be obtained. However, in that case the instantaneous current in the loop ($I_P^I$) at the time of cut-off is non-zero and the energy stored on the inductor $\left(\frac{1}{2}L_2\left(I_P^I\right)^2\right)$ is wasted. To recover this stored inductor energy, a diode ($D_F$) is added to the circuit as shown in Fig. S2(b) that gives path to freewheeling inductor current and in turn continue to charge the TENG capacitor. As shown in the waveform plots of Fig. S2(c), the switch on duration of $S_2$ is adjusted to obtain the required $V_{pb}^{II}$ value. While, the switch $S_4$ is enabled for greater than or equal to half the $L_2 - C_{T,min}$ resonator cycle for completing the freewheeling current loop and is automatically cut-off as the inductor is completely discharged.

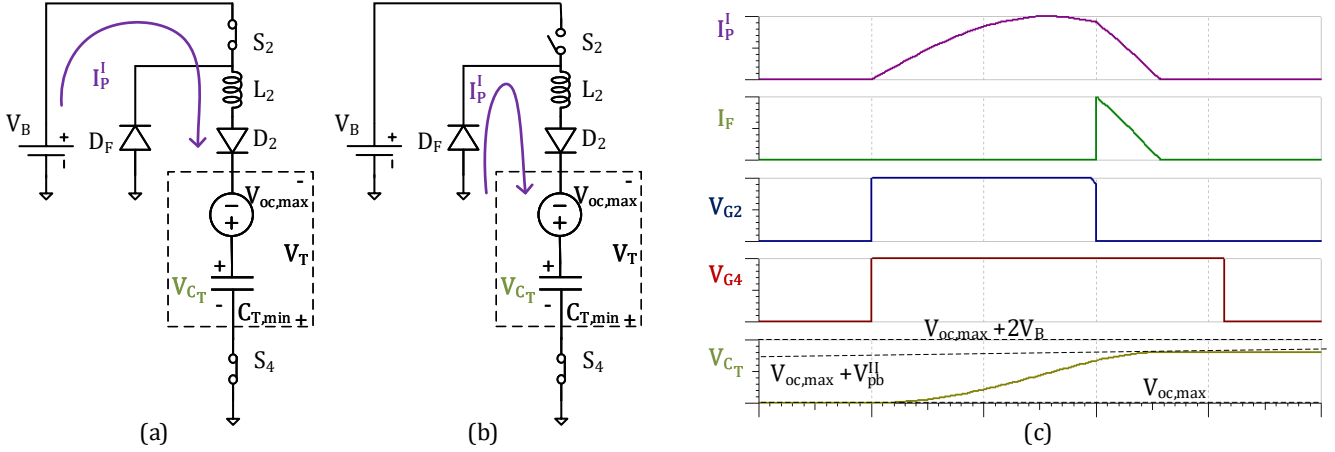

Fig. S2. Simplified circuit during State II++ (a) with switches $S_2$ and $S_4$ closed. (b) with switch $S_2$ open and $S_4$ closed enabling the inductor current flow through the freewheeling diode ($D_F$). (c) Simulated waveforms for the pre-biasing loop current ($I_P^I$), freewheeling current ($I_F$), enable signals for switches $S_2$ ($V_{G2}$) and $S_4$ ($V_{G4}$), and TENG capacitor voltage ($V_{C_T}$).

## S.VI. Energy output comparison of SCE versus pSCE keeping identical volume of inductors

The pSCE circuit uses 4 inductors as opposed to 2 used by SCE. It is known that quality factor of inductors is proportional to its volume. Then by choosing inductors of half the quality factor as that used in SCE, the pSCE output can be compared to that of SCE at a constant magnetics volume [3], [4].

With the quality factors, $Q_f^I$ and $Q_f^{II}$ and their normalized forms, $\alpha^I$ and $\alpha^{II}$ defined in Eq. (S1) and Eq. (S5), respectively, the per-cycle energy output of the SCE circuit is given by Eq. (S6) and restated below:

$$E_{SCE} = E_{SCE}^I + E_{SCE}^{II} = \frac{1}{2}\left(\frac{\alpha^I}{\beta} + \alpha^{II}\right)C_{T,min}V_{oc,max}^2. \tag{S14}$$

Then to maintain equal inductor volume, the quality factors of the pSCE circuits shall be taken as $Q_f^{'I} = \frac{Q_f^I}{2}$ and $Q_f^{'II} = \frac{Q_f^{II}}{2}$ at States I and II, respectively. The normalized quality factors of the pSCE circuit can then be found as,

$$\alpha^{'I} := e^{\frac{-\pi}{2Q_f^{'I}}} = e^{\frac{-2\pi}{2Q_f^I}} = (\alpha^I)^2. \quad \text{Similarly, } \alpha^{'II} = (\alpha^{II})^2. \tag{S15}$$

We need to consider the resistive loss due to the limited quality factor of inductors during both extraction and pre-biasing steps in the pSCE circuit. For the case where TENG is pre-biased for half the LC oscillation cycle to the maximum achievable value using load battery of voltage $V_B$ at both States I and II, based on Eq. (S11), the net output of the pSCE circuit can be derived as,

$$E_{pSCE,net} = \frac{1}{2}C_{T,min}\Bigg[\left(\alpha^{II}\left(V_{oc,max} + \left(1+(\alpha^I)^2\right)\beta V_B\right)\right)^2 + \frac{\left(\alpha^I\left(V_{oc,max} + \left(1+(\alpha^{II})^2\right)V_B\right)\right)^2}{\beta}$$
$$- 2\left(1+(\alpha^{II})^2 + \left(1+(\alpha^I)^2\beta\right)\right)V_B^2\Bigg]. \tag{S16}$$

Using the above derived equations (Eq. (S14) and Eq. (S16)) and the implemented TENG's parameters (Table I), we co-plot the per-cycle outputs of SCE and pSCE as a function of SCE circuit's quality factor at State I ($Q_f^I$) at two different battery loads of 5V and 15V in Fig. S3. We use the approximate relation, $Q_f^{II} \approx \sqrt{\bar{\beta}} Q_f^I$ between the quality factors at State I and II. As can be seen from Fig. S3, even with half the quality factor inductors, pSCE provides higher energy output than SCE using inductors of quality factor as low as 4.

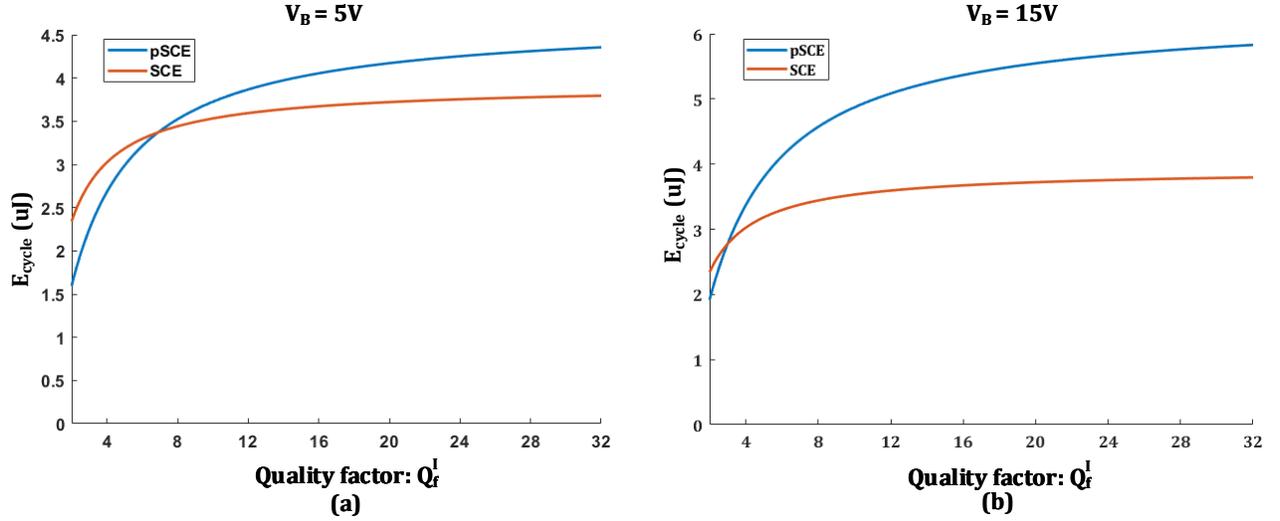

Fig. S3. Per-cycle energy output of pSCE and SCE circuits against quality factor at identical inductor volume for battery load of (a) 5V and (b) 15V.

## S.VII. MEASUREMENT OF DYNAMIC TENG CAPACITANCE

TENG Capacitance ($C_T(t)$) is measured using the method based on phase response presented in [5], [6]. Below figure, shows measured $C_T$ for our experimental TENG over 5 cycles of operation. Maximum and minimum TENG capacitance ($C_{T,max}$ and $C_{T,min}$) are deduced from the below plot by first taking a moving average with span of 2% and then finding the maxima and minima. Values obtained are $239.23pF$ and $75.97pF$, respectively.

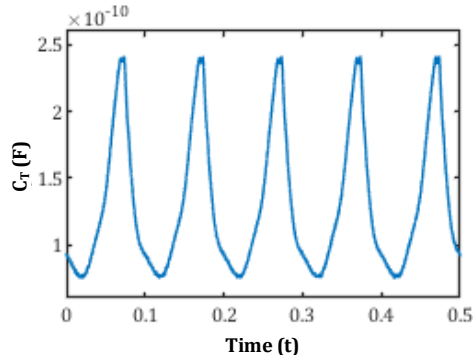

Fig. S4. Measured time varying TENG capacitance ($C_T(t)$)


    }
\fi

\ifCLASSOPTIONcaptionsoff
 \newpage
\fi

\end{document}